\documentclass[journal]{IEEEtran}
\usepackage{amsmath,amsfonts}
\usepackage{array}
\usepackage{tikz}
\usepackage{algorithm}
\usepackage{algpseudocode}
\usepackage{pgfplots}
\usepackage{physics}
\usetikzlibrary{arrows.meta,calc,shapes.geometric}%
\usepackage{standalone}    %
\usepackage{amsmath,amssymb,amsfonts}
\usepackage{graphicx} 

\usepackage{booktabs}
\usepackage{multirow}
\usepackage{siunitx}
\usepackage{adjustbox}
\usepackage{graphicx}
\usepackage[caption=false,font=normalsize,labelfont=sf,textfont=sf]{subfig}
\usepackage{textcomp}
\usepackage{stfloats}
\usepackage{url}
\usepackage{verbatim}
\usepackage[none]{hyphenat}
\usepackage{mathtools}
\usepackage{cite}

\usepackage{xcolor}
\definecolor{panelgreen}{RGB}{170,214,126}

\usepackage{graphicx}

\providecommand*{\M}[1]{\mathbf#1}        
\providecommand*{\V}[1]{\boldsymbol#1} 
\providecommand*{\UV}[1]{\hat{\boldsymbol#1}}  
\providecommand*{\T}[1]{\mathrm{#1}}  
\providecommand*{\eu}{\ensuremath{\T{e}}}  
\providecommand*{\ju}{\ensuremath{\T{j}}} 
\providecommand*{\diff}{\operatorname{d}\!}  

\begin{document}

\title{Properties of Near-Field Focusing for Three-Dimensional Large Intelligent Surface}

\author{Jiawang Li, Mats Gustafsson, Alireza Saberkari, Buon Kiong Lau

\thanks{Manuscript received \today. This work was supported in part by a Project Grant in ELLIIT 
Call  D,  and  in  part  by  NextG2Com  (grant  no.  2023-00541) 
funded by the VINNOVA program for Advanced Digitalisation.  (Corresponding author: \textit{Jiawang Li}).}%
\thanks{Jiawang Li, Mats Gustafsson and Buon Kiong Lau are with the Department of Electrical and Information Technology, Lund University, 22100 Lund, Sweden (e-mail: {\{jiawang.li, mats.gustafsson, buon\_kiong.lau\}}@eit.lth.se).

Alireza Saberkari is
with the Department of Electrical Engineering, Linköping
University, 58183 Linköping, Sweden. (e-mail: alireza.saberkari@liu.se).
}%
}

\markboth{Properties of Near Field Focusing for Three-Dimensional Large Intelligent Surface, \today}%
{Li: Properties of Near Field Focusing for Three-Dimensional Large Intelligent Surface}

\maketitle

\begin{abstract}
This work investigates near-field focusing using a three-dimensional (3D) large intelligent surface (LIS) across frequencies and polarizations. Specifically, the LIS elements are distributed in 3D space within a long corridor, rather than being confined to a single planar aperture, and the focal point is located at a prescribed position in the radiating near field. By formulating optimization problems under both local and global power constraints, we obtain the corresponding optima. For continuous apertures, the optimal current magnitude distribution matches time-reversal (TR) solution under the global constraint and conjugate-phase (CP) solution when the local constraint dominates. When both constraints are active, the solution assigns larger excitation magnitudes to elements closer to the illumination field. This behavior remains invariant with respect to frequency and polarization for a fixed-size LIS. These findings are consistent to the more practical case of using discretized apertures in the form of Hertzian dipole arrays, studied using both analytical results and full-wave simulation. In addition, with the CP method, specific polarizations lead to identical transverse and longitudinal resolution, in contrast, under the TR method, these quantities can differ across polarizations.
\end{abstract}

\begin{IEEEkeywords}
Large intelligent surface (LIS),
near field focusing, time reversal,
conjugate phase,
Hertzian dipole.
\end{IEEEkeywords}

\section{Introduction}
\begin{figure}[t]
    \centering
    \includegraphics[width=0.6\linewidth]{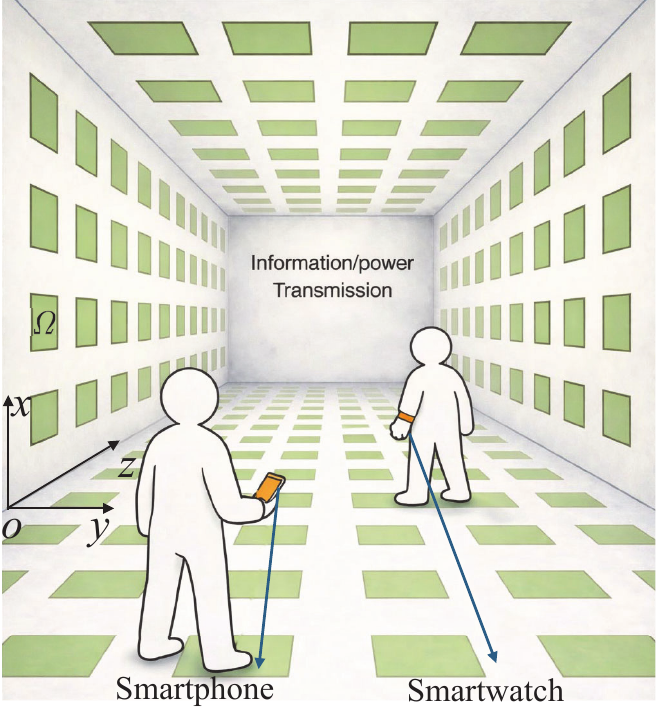}
    \caption{LIS application scenarios. The figure illustrates users moving within a long corridor. The corridor is surrounded by LIS, with each LIS panel represented by a region $\it{\Omega}$. The longitudinal polarization is along the corridor axis (\(z\)-directed), whereas the transverse polarization lies across the corridor (\(x\)- or \(y\)-directed), with respect to the coordinate system shown.}
    \label{fig:LIS_scenario}
\end{figure}

\IEEEPARstart{L}{arge} intelligent surface (LIS) is a promising concept for meeting the stringent requirements of 6G in sub-10 GHz bands \cite{hu2018beyond1,hu2023design}, given the constraints on frequency spectrum resources. With the integration of electronics and wireless connectivity into future man-made structures, LIS enables environments to become electronically active. As depicted in Fig.~\ref{fig:LIS_scenario}, 
LISs can comprise numerous panels that are installed on walls, ceilings, and even floors, and each panel can feature tens or hundreds of antenna elements. These elements with proper weighting can focus (or beamform) signal power. For instance, they can communicate with smartphones while simultaneously charging devices such as smartwatches. In this scenario, the electromagnetic waves generated by the LIS beamforming are no longer far-field plane waves \cite{schmidt2008fully} but near-field spherical waves\cite{bosma2021near, broquetas1998spherical, garnica2013wireless}. Consequently, conventional far-field beamforming methods \cite{van1988beamforming, zaharis2020effective} are no longer valid, prompting a growing focus on near-field methods.

Current research on near-field beamforming techniques primarily aims at maximizing the electric field within a specific region under certain prescribed conditions \cite{nepa2017near, buffi2012design, kosasih2024finite, fink2002time, nguyen2006time, cassereau2002time, huang2020synthesis, huang2018near, chen2025structured, gonzalez2019design, alvarez2014near}. Beamforming techniques can be broadly categorized into two types. The first type is based on the conjugate-phase (CP) method \cite{nepa2017near, buffi2012design, kosasih2024finite} and the time reversal (TR) method \cite{fink2002time, nguyen2006time, cassereau2002time}, whereas the second one employs various optimization algorithms \cite{huang2020synthesis, huang2018near, chen2025structured, gonzalez2019design, alvarez2014near}. 

The focal point and field characteristics of rectangular planar arrays have been analyzed in  \cite{nepa2017near, buffi2012design, kosasih2024finite}. A theoretical model of closed TR cavities is developed to understand and optimize ultrasonic wave focusing in homogeneous and weakly inhomogeneous media \cite{cassereau2002time}. However, these studies consider one or two-dimensional arrays and do not take into account the impact of antenna polarization.
Some optimization methodologies, such as convex optimization \cite{ huang2020synthesis}, compressive sensing (CS) \cite{huang2018near}, structured sparsity learning (SSL)-based deep neural
network (DNN) \cite{chen2025structured}, and Levenberg-Marquardt algorithm \cite{gonzalez2019design} had been utilized. Other design approaches also exist, including methods that achieve near-field focusing solely through optimization of the element phases \cite{alvarez2014near}. Most of these methods rely on different optimization strategies to determine the element excitation coefficients under various design conditions, but they do not explicitly consider the implications of different power constraints in practical implementations.

The investigated LIS scenario in this work differs substantially from the prior studies. First, most existing large-scale array applications assume a base-station deployment where users are located far from the array\cite{kosasih2024finite}, and thus the received signal amplitude from each array antenna can be treated as approximately constant. In contrast, in the LIS configuration, the antenna elements are much closer to the user, and the corresponding amplitude variations can no longer be treated as uniform. Second, compared with conventional near-field focusing studies, the LIS concept envisions an extremely large number of array elements rather than small to moderate-size arrays \cite{huang2020synthesis}. Third, prior designs predominantly focus on two-dimensional (2D) planar geometries, whereas a LIS three-dimensional (3D) array can surround users, yielding more distinctive focusing characteristics.

In this paper, we investigate CP and TR under a unified optimization framework.
Both can be viewed as solutions to the same focusing problem, but under different amplitude constraints.
Specifically, CP enforces a uniform-amplitude excitation\cite{nepa2017near, buffi2012design, kosasih2024finite}, whereas TR yields a tapered amplitude distribution proportional to the illuminating field.
This contrasts with some works in other fields that regard the two as essentially the same method ~\cite{katko2012phase, ohno2012ultrasonic, chabalko2016electromagnetic}. A key contribution is treating polarization as an explicit design degree of freedom. Most existing near-field focusing/shaping studies assume single-polarization~\cite{nepa2017near, buffi2012design, kosasih2024finite, fink2002time, nguyen2006time, cassereau2002time, huang2020synthesis, huang2018near, chen2025structured, gonzalez2019design, alvarez2014near}, and may overlook practical limitations; related results in vector-field focusing also suggest that polarization-aware synthesis can reshape the 3D focal region~\cite{chen2006three}. We further study both continuous-aperture and discretized-element implementations, and derive theoretical resolution limits as fundamental performance benchmarks. The main findings are summarized as follows.

For both continuous aperture characterized by current density and discrete array, the optimal amplitude for a fixed-size aperture is frequency-invariant yet polarization-dependent. Under a local power constraint, the optimal amplitude coincides with the CP solution, whereas under a global power constraint, it aligns with the TR solution. We also derive a closed-form vector expression for the focal electric-field distribution when the receiving antenna is a Hertzian dipole.

The focal-point intensities produced by different polarizations remain stable inside the corridor, in contrast to conventional planar arrays \cite{nepa2017near, buffi2012design, kosasih2024finite} where amplitude loss occurs as the focal point deviates from the central axis. For a longitudinally polarized Hertzian dipole array (see Fig.~\ref{fig:LIS_scenario}), when scanning along the array axis, TR yields nearly constant orthogonal field components whereas CP remains strongly distance dependent. Moreover, CP achieves equal transverse and longitudinal resolution in this configuration, a property not observed under TR.

The rest of the paper is organized as follows. Section~II formulates the optimization problem for the LIS array. In Section~III we analyze the aperture using a continuous current distribution, compare the current amplitude profiles and focal characteristics for different methods and polarizations. Section~IV studies the more practical case of using Hertzian dipoles arrays instead of continuous currents, with similar trends obtained. In Section~V we derive closed-form expressions that further elucidate the unique focusing-field and resolution properties of 3D LISs. Section~VI concludes the paper.

\textit{Notation:} Throughout this paper, boldface letters indicate vectors
and boldface uppercase letters designate matrices. Superscripts
$(\cdot)^{-1}$, $(\cdot)^{*}$, $(\cdot)^{\T{T}}$ and $(\cdot)^{\T{H}}$ stand for the inverse,
complex conjugate, transpose and Hermitian transpose, respectively. In addition,
$\T{Re}\{\cdot\}$ takes the real part.
\section{Power-Constrained Optimization Formulation}

The beamforming weight of each element in a LIS is constrained by several factors, including per-channel hardware limits on the maximum output power, a constraint on the total system power, and regulatory limits on the maximum radiated power density (typically \(10~\mathrm{W/m^2}\) \cite{recommendation1999limitation}).
 Therefore, in this work, two types of power constraints are considered: local power and global power constraints. The local power constraint, which refers to the maximum output power of each radio frequency chain, is typically limited by the 1-dB compression point of the final stage power amplifier (PA) \cite{raab2002power}. The global power constraint requires that the total transmit power of the entire LIS system remain below a prescribed limit, which should comply with the applicable spectral mask. We further introduce a hybrid constraint that simultaneously incorporates both local and global power limitations. 

We adopt a beamforming strategy that maximizes the power at the focal point \(\V r_\T{f}\),  directly determined by the electric field $E$ for a given polarization, subject to the power constraints. Accordingly, the optimization problem is in the form 
\begin{equation}\label{eq:local_tptal_power_constraints}
\begin{aligned}
 & {\T{maximize}} && E\ \text{at focal point}\\
& \T{subject}\ \T{to} && \text{Local power constraint}\\
& && \text{Total power constraint}
\end{aligned}
\end{equation}

The two aforementioned power constraints are relevant for an LIS panel comprising numerous antenna elements. The total power $P_0$ is the net power supplied to the LIS panel without considering the loss from the feeding network, equal to the sum of the powers of all elements. For each antenna element, we define a local power controlled by its complex excitation weight $w_n$ in the vector $\M w = [w_1,\ldots,w_N]^\T{T}$, where \(N\) is the total number of elements.

Accordingly, maximizing the electric field intensity at a prescribed spatial point and polarization can now be explicitly posed as an optimization problem subject to both a global power budget $P_0$ and local power constraints
\begin{equation}\label{eq:local_tptal_power_constraints_equation}
\begin{aligned}
& {\T{maximize}} && \left|E (\V r_\T{f},\UV e)\right| \\
& \T{subject}\ \T{to} &&
 |w_n|\;\le\; w_{\T m},\quad n=1,\dots,N\\
& && \M{w}^{\T H}\M{R}\,\M{w}\;\le\;2P_0,
\end{aligned}
\end{equation}
 where $\UV e $ denotes the polarization unit vector at focal point, thus $E(\V r_\T{f},\UV e)=\V E(\V r_\T{f}) \cdot \UV e$ is the scalar component of the electric field $\V E$ along $\UV e$ at the focal point. \(w_{\T m}\) denotes the upper bound of the \(w_n\).  \(\M R\) denotes the port-resistance matrix defined by the source of the LIS panel. Neglecting mutual coupling renders \(\M R\) diagonal of size \(N\times N\). Assuming the LIS is lossless, this implies that \(P_0\) equals the actual radiated power of the entire panel.
 When coupling is included, one can compensate by pre-multiplying with the array port-impedance matrix \cite{hui2007decoupling}. If we interpret $w_n$ directly as port currents $I_n$ and assume each port sees an equivalent resistance $R_0$, then in a simple circuit model the total transmitted power is \(P_0 = R_0/2\sum_n|I_n|^2\). For example, with \(P_0=1\,\T{W}\), \(R_0=50\,\Omega\) and \(N=2000\), the maximum current is \( I_{\max} = \sqrt{2P_0/R_0/N} \approx 0.0045\,\T{A} \), hence \(w_{\T m}\approx0.0045\,\T{A}\). For consistency, all subsequent simulations assume a $50~\Omega$ port impedance and the unit of \(w_\T{m}\) is Ampere.

 We impose a local maximum amplitude constraint ($|w_n|\le w_{\T m}$) instead of a local power constraint.
 This choice is consistent with common element-driving schemes: antenna elements are typically excited by either a voltage source or a current source, and any local power constraint in the CP formulation can be equivalently represented as an amplitude bound through the source impedance.  

Consider an $N$-element array, for the $n$-th element, let
$\M{h}_n \in \mathbb{C}^3$ denote the vector channel
coefficient from that element to the focal point, whose entries represent the complex field components along polarizations at the focal location. Stacking these coefficients gives the
channel representation
$\M{h} = [\M{h}_1^{\T T}, \ldots, \M{h}_N^{\T T}]^{\T T}$, where
$\M{h}_n$ is a vector-valued coefficient collecting all
orthogonal polarization components at the focal point, so that
$\M{h}$ implicitly represents the full dyadic
channel between the array and the focal point.

Under these local and global power constraints, when the array operates in the per-element limited regime the total power budget is effectively nonbinding, and the optimal solution reduces to the CP excitation.
Under the per–element amplitude constraint, we can get the optimal weights and the resulting focal electric field \(E_{\T{CP}}\)
\begin{equation}
  w_n
  = w_{\T m}
    \frac{(\UV{p}_n\cdot\M{h}_n\cdot\UV{e})^\ast}
         {\bigl|\UV{p}_n\cdot\M{h}_n\cdot\UV{e}\bigr|},
  \qquad
  E_{\T{CP}}
  = w_{\T m}\sum_{n=1}^{N}
    \bigl|\UV{p}_n\cdot \M{h}_n \cdot \UV{e} \bigr|,
  \label{eq:Ecp}
\end{equation}
where $\UV p_n$ is the polarization direction of the \(n\)th transmitting antenna element. This regime applies when $P_0$ is sufficiently large so that the per–element constraint dominates. All elements transmit with equal amplitude and apply phase precompensation to cancel the propagation phase, so that all paths add in phase at the focus, the focusing field is therefore governed primarily by the sum of the path magnitudes. In scenarios with multiple polarization components, if one component dominates the others, the remaining components contribute negligibly and can be treated as noise.

The electric field at the focal point is a vector quantity, whereas a single-polarized receive antenna only responds to the component along its polarization unit vector $\UV e$. Consequently, \eqref{eq:local_tptal_power_constraints}--\eqref{eq:local_tptal_power_constraints_equation} can be interpreted as maximizing an equivalent scalar channel obtained by projecting $\V h_n$ onto $\UV e$. Under a per-element amplitude constraint, the optimal solution reduces to the CP weights in \eqref{eq:Ecp}, where each element phase is aligned with the projected channel to form the focal field. The total power constraint in \eqref{eq:local_tptal_power_constraints_equation} is then enforced by a single common scaling of these CP weights: it is inactive when the per-element constraint dominates, and otherwise uniformly scales down the amplitudes to satisfy the total power budget.

In the global power limited regime, the per–element limit is left out and the optimum coincides with the TR solution, where the field at the focus is measured, time-reversed, and re-emitted so that propagation reconstructs the target phase. With a global power budget $P_0$, the optimal weights \(\M w\) and focal field \(E_{\T{TR}}\) are
\begin{equation}\label{eq:tr_weight and e-field}
\M w = \sqrt{2P_0}\,\frac{\M R^{-1}\M h^{*}}
              {\sqrt{\M h^{\T H}\M R^{-1}\M h}},\qquad
E_{\T{TR}} = \sqrt{2P_0\,\M h^{\T H}\M R^{-1}\M h}.
\end{equation}
This regime applies when $w_{\T m}$ is sufficiently loose in \eqref{eq:local_tptal_power_constraints_equation} so that only the total power matters. The \(\M h\) here accounts for the different polarization components at each port, and is therefore slightly different from the \(\M h\) defined earlier.

The two methods are similar in that both assign the phase of each element by phase conjugation. They differ in their amplitude allocation, the CP method uses uniform-amplitude excitation, whereas the TR scheme assigns larger amplitudes to elements that are more closer to the focal point. For a fixed total power budget, increasing the number of elements reduces the power available to each element under CP. Once the resulting per-element power becomes comparable to the noise level, CP effectively breaks down. By contrast, TR concentrates power on a subset of elements and assigns only negligible power to elements far from the focus. Under these conditions, the advantage of TR becomes more pronounced.

\section{Electric Field From Continuous Currents On A Cylinder}
As the radiation of an LIS is determined by its surface current distribution, the radiated electric field can be modeled as a linear functional of the surface current. By the equivalence principle \cite{kong1975theory}, any physically realizable exterior field can be synthesized by suitable electric and magnetic surface current densities on a LIS panel region $\it\Omega$ in Fig.~\ref{fig:LIS_scenario}. Therefore, we formulate the design as optimizing electric and magnetic surface current densities over $\it\Omega$ so that propagation yields the desired field in the target region, that is $E (\V r,\UV e) = \T {\T{Re}}\bigl (\M h \M w\bigr)$. In this case, $\M h$ denotes the dyadic Green’s-function matrix for the continuous current distribution, with the same physical interpretation as channel vector $\M h$ (see Appendix~\ref{app:proof_A}), $\M {w}$ is the complex weight (equivalently the excitation coefficients for basis function in the Method of Moments (MoM) \cite{harrington1993field}). The radiated field can be expressed as \cite{eibert2015electromagnetic}
\begin{equation}\label{eq:e-field_expression}
\V{E}(\V{r}) = \int_{\it{\Omega}} 
{{\V{G}}}_\T{J}^\T{E}(\V{r}, \V{r}') \cdot \V{J}(\V{r}') 
+ {{\V{G}}}_\T{M}^\T{E}(\V{r}, \V{r}') \cdot \V{M}(\V{r}')  \diff \T S',
\end{equation}
where $\V{J}(\V {r}')$ and $\V{M}(\V{r}')$ are the electric and magnetic current densities, and 
${{\V{G}}}_\T{J}^\T{E}$ and ${{\V{G}}}_\T{M}^\T{E}$ 
are the corresponding dyadic Green’s functions, respectively. $\V{r}$ and $\V{r}'$ denote the position vectors of the field point and the source point.
The dyadic Green's functions are given by \cite{eibert2015electromagnetic}
\begin{equation}\label{eq:electric current_field}
\V{G}_\T{J}^\T{E}(\V{r}, \V{r}')
= -\,\ju {k \eta_0}
\Bigl(\M{I}_3 + \frac{1}{k^2}\,\nabla\nabla\Bigr)
\frac{\eu^{-\ju k\lvert\V{r}- \V{r}'\rvert}}
     {4\pi\lvert\V{r}- \V{r}'\rvert}\,,
\end{equation}
\begin{equation}\label{eq:magnetic current_field}
\V{G}_\T{M}^\T{E}(\V{r}, \V{r}')
= -\,\nabla\times\M{I}_3\,
\frac{\eu^{-\ju k\lvert\V{r}- \V{r}'\rvert}}
     {4\pi\lvert\V{r}- \V{r}'\rvert}\,,
\end{equation}
where $k$ is the free space wavenumber and \(\eta_0\) is the wave impedance in the free space and ${\M{I_3}}$ is the unit dyad. The electric‐current source and the magnetic‐current source each represent a different class of antenna element: the former can model dipole or patch antennas, while the latter corresponds to slot‐type antennas \cite{balanis2016antenna}.

\begin{figure}[t]
    \centering   \includegraphics[width=0.7\linewidth]{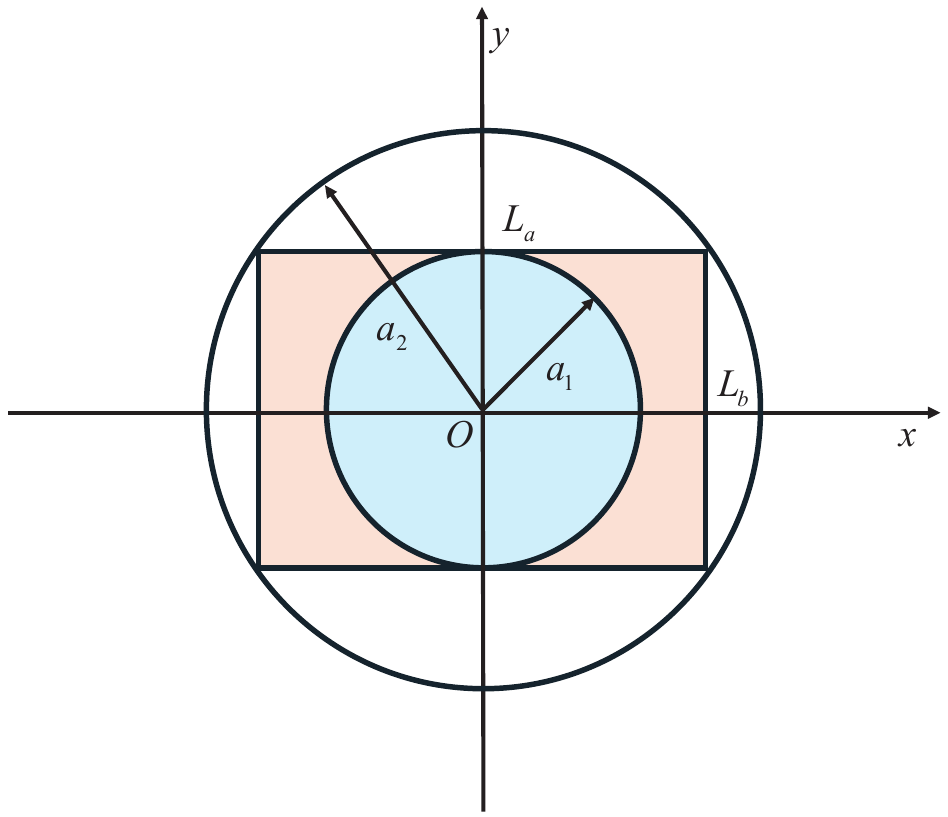} 
    \caption{Indoor LIS scenario of a long corridor as in Fig.~\ref{fig:LIS_scenario}. The cross-sectional transformation replaces the original rectangular cross-section with an inscribed circle of radius \(a_1\) and a circumscribed circle of radius \(a_2\). In this model, the $z$-axis is aligned with the corridor's length.}
    \label{fig:LIS application scenarios indoors}
\end{figure}

For most elongated rooms or corridors, the height and width are often different, this discrepancy complicates the analysis of near-field focusing characteristics. 
Therefore, to reduce the complexity introduced by having both width and height as parameters, we replace the rectangular cross-section with two circles of different sizes, as shown in Fig.~\ref{fig:LIS application scenarios indoors}.
Specifically, if the corridor or elongated room has width \(L_a\) and height \(L_b\), we use the inscribed circle of the shorter side with the radius \(a_1\) (i.e., \(2a_1 = L_b\)) and the circumscribed circle of the longer side with the radius \(a_2\)  in place of the rectangle. The corridor depth refers to the longitudinal extent along the corridor axis (the $z$ direction), denoted by $L$.

This approach offers three key advantages: (1) When a rectangle has unequal length and width, its focal characteristics are not the same along the two axes (i.e., \(x\) and \(y\)-axes) and would otherwise require repeated case-by-case analysis; (2) The rectangle’s surface area lies between those of two bounding circular apertures with radii chosen to upper and lower bound the area, ensuring that the achievable peak electric field is bounded within the interval defined by the two circular cases; (3) By replacing the two side lengths with a single radius parameter, the model reduces to one degree of freedom, which facilitates closed-form expressions for near-field quantities.

\begin{figure}[t]
    \centering
    \includegraphics[width=0.7\linewidth]{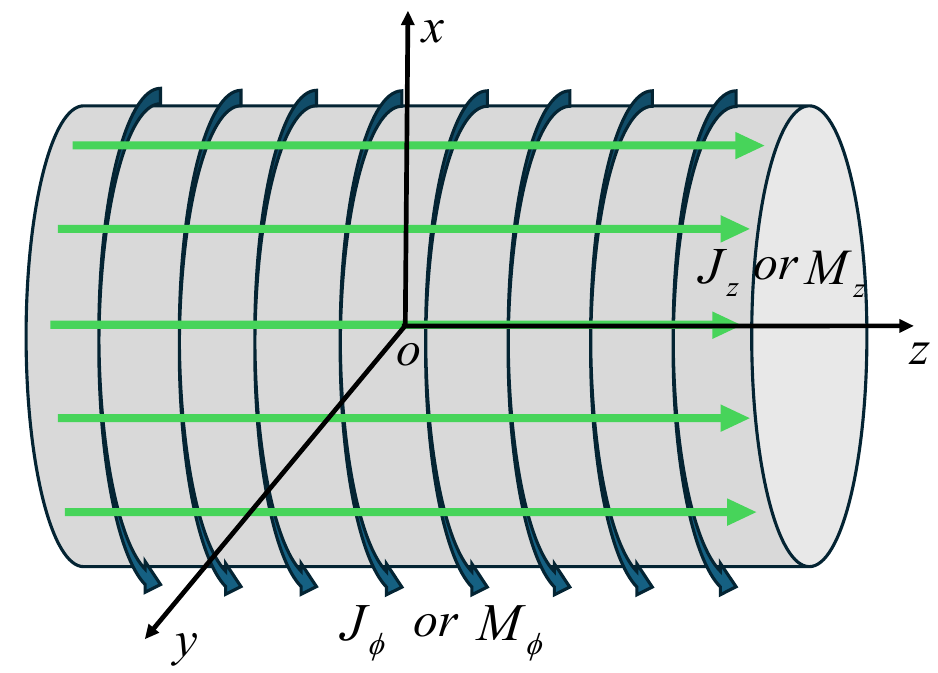} 
\caption{Decomposition of the surface current on a cylinder either electric or magnetic into its axial component \(J_{\T z}\) or \(M_{\T z}\) and its circumferential component \(J_{\T \phi}\) or \(M_{\T \phi}\).}
    \label{fig:Decomposition}
\end{figure}

As shown in Fig.~\ref{fig:Decomposition},
the surface electric current on the cylinder can be decomposed into an azimuthal component and an axial component \cite{mautz1977h}, i.e.,
\begin{equation}
\V{J}(\phi,z)
= J_{\T \phi}(\phi,z)\,\UV \phi
+ J_{\T z}(\phi,z)\,\UV z,
\end{equation}
where \(J_{\T \phi}\) and \(J_{\T z}\) denote the circumferential and axial current densities, respectively. Similarly, the surface magnetic current admits the analogous decomposition
\begin{equation}
\V{M}(\phi,z)
= M_{\T \phi}(\phi,z)\,\UV \phi
+ M_{\T z}(\phi,z)\,\UV z,
\end{equation}
where \(M_{\T \phi}\) and \(M_{\T z}\) representing the azimuthal and axial magnetic‐current densities, respectively. These two current components can likewise be interpreted as antennas with orthogonal polarizations, effectively realizing a dual-polarized antenna for the given LIS distribution.
Since the electric current and magnetic current sources are independent, we can first analyze one source under various power‐constraint scenarios and then compare the fields they produce to highlight their similarities and differences. To facilitate the analysis, we expand the $\phi$-dependence using a Fourier series with angular harmonics $\eu^{ \ju \nu\phi}$, $\nu\in\mathbb{Z}$ denotes the azimuthal Fourier-mode index.

 The problem \eqref{eq:local_tptal_power_constraints_equation} can be solved using convex optimization techniques 
such as the MOSEK solver \cite{andersen2000mosek}. We also provide a simpler algorithm (Algorithm~1, see Appendix \ref{app:Algorithm 1})
that achieves the same solution as the MOSEK solver for \eqref{eq:local_tptal_power_constraints_equation}, by setting each element’s weight to \(w_\T {m}\) scaled by the magnitude of its Green-function entry to the focal point while using the CP. The algorithm delivers the optimal amplitude distribution efficiently and quickly. In the MoM-based simulations, the axial and azimuthal directions are discretized using 2000 and 360 segments, respectively.

\begin{figure}[t]
  \centering
    \centering    \includegraphics[width=0.9\linewidth]{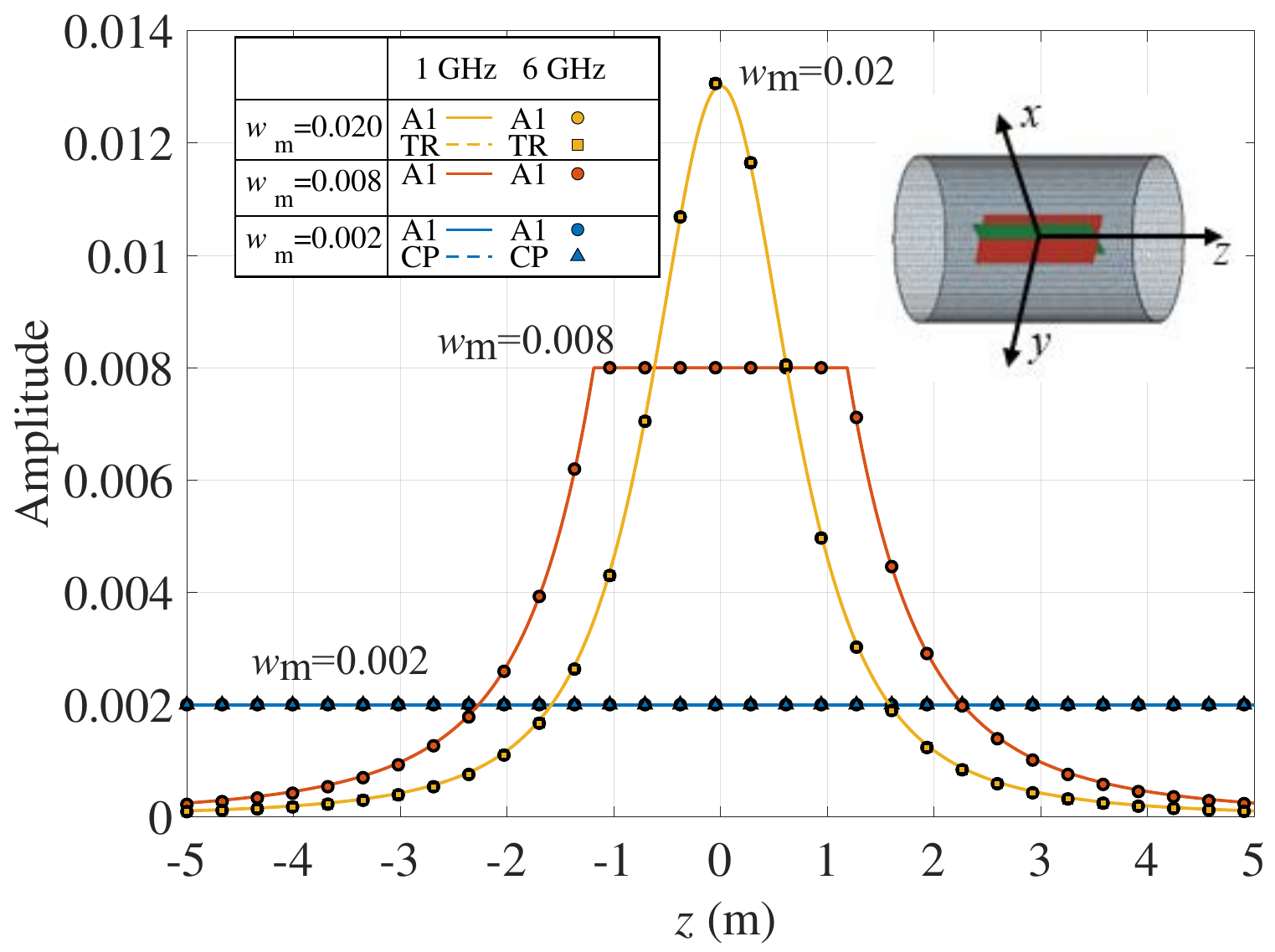}
  \caption{Amplitude distribution of $\M w$ along the longitudinal direction of a cylindrical surface
of radius 1 m and length 10 m when \(\UV e = \hat{\V z}\). \(P_0 = 1~{\T W}\). The focal point optimized lies at $\V r_\T{f} = \V 0$ with \(z\)-polarization. \(\T{A1}\) denotes Algorithm 1.}
\label{fig:global_opt_z}
\end{figure}

Figure \ref{fig:global_opt_z}
shows the current amplitude distributions for a \(10\,\T{m}\)-long cylindrical corridor of radius \(1\,\T{m}\) when \(\UV e= \hat{\V z}\). Phases are omitted because the optimum equals the conjugate phase. For a fixed-size aperture, the amplitude distributions at 1\,GHz and 6\,GHz are same. When local power (\(0.2  \ \T W, w_{\T m} = 0.002\, \T A/\T m\)) is much less than total power (1 W), the optimal amplitude coincides with the CP solution \eqref{eq:Ecp}. When local power (\(20 \ \T W, w_{\T m} = 0.02\, \T A/\T m\)) is much larger than total power, it coincides with the TR solution \eqref{eq:tr_weight and e-field}. When local power (\(3.2 \ \T W, w_{\T m} = 0.008\, \T A/\T m\)) is only slightly more than total power, an intermediate pattern appears in which elements nearest the focus reach the local limit first and the remaining elements follow a TR-like distribution.

When the focal point lies on the z-axis, only the azimuthal mode \(m=0\) contributes to the on-axis \(z\)-polarized field, and all other modes do not contribute. Hence, it suffices to retain the \(m=0\) mode. For the subsequently synthesized \(x\)-polarized electric field, if the focal point is on the axis, only the \(m=\pm 1\) modes need to be kept. For off-axis focusing, additional modes contribute. Under dominant local constraints, maximizing the focal field requires progressively incorporating higher order azimuthal modes until the peak value converges. Moreover, the larger the radial offset of the focus, the greater the number and order of azimuthal modes required. Use a sufficiently large azimuthal-mode truncation order to ensure convergence of the CP electric-field calculation.

\begin{figure}[t]
  \centering  \includegraphics[width=1\linewidth]{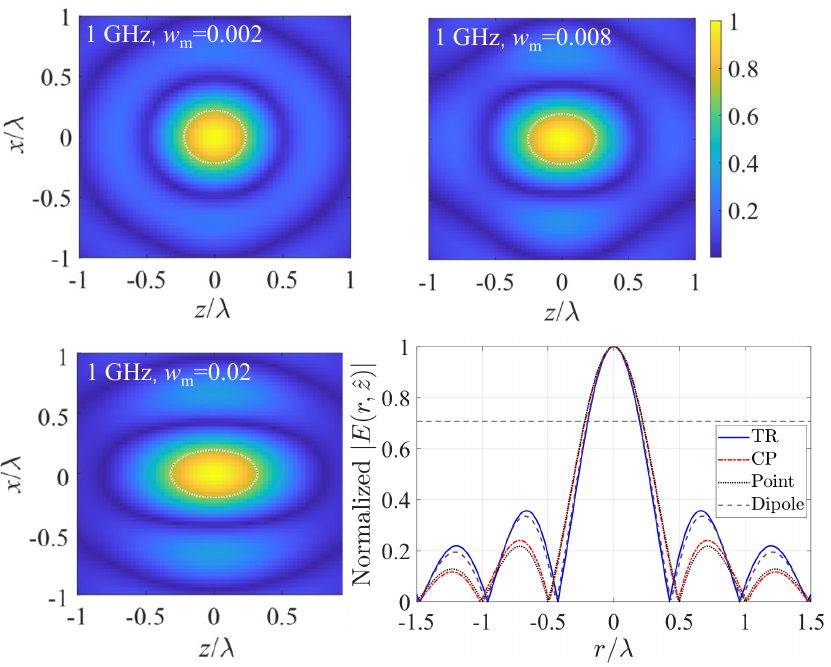}
  \caption{Normalized co-polarized electric field \(E_{\T z}\) optimization result for a cylindrical surface of radius 1 \(\T{m}\) and length 10 \(\T{m}\) as well as the normalized focusing field for point-source \eqref{kernel} and Hertzian-dipole \eqref{kernel_vector}, including the 1D normalized focal gain along the $z$-axis. The focal point lies at \(\V r_\T{f}=\V 0\) with \(z\)-polarization. The 2D cut figure is along the \(x\) axis and is symmetric with respect to both $x$ and $z$. ``Point'' follows \eqref{kernel}, while ``Dipole'' follows \eqref{kernel_vector} for $\vartheta = 90^\circ$. The 3-dB boundary (white ring) corresponds to the half-power region, equivalently, it is the \(1/\sqrt{2}\)-amplitude contour shown as the dashed line in the 2D cut.}
  \label{fig:z_polarzation_16}
\end{figure}

Figure \ref{fig:z_polarzation_16}
shows the electric field distribution in the focal region on the \(xz\)-plane at $1\,\T{GHz}$ and $6\,\T{GHz}$ (for $6\,\T{GHz}$, the results behave similarly to those at $1\,\T{GHz}$, the focal-spot size scales proportionally with the wavelength, and the corresponding results are therefore omitted). Because the \(xz\) and \(yz\)-planes are equivalent by symmetry, only the \(xz\)-plane results are shown. As the design evolves from the CP to the TR method, the focal spot becomes increasingly asymmetric. The focal-spot size scales with the wavelength, while the fields exhibit consistent spatial distributions across frequencies. 

In fact, the field distribution obtained along the $x$-axis cut closely
matches the TR distribution \(\tilde{K}(\V{r})\) for a scalar point source in a closed cavity~\cite{nguyen2006time}
\begin{equation}
\tilde{K}({r})
= \frac{\sin(kr)}{kr}
={\operatorname{sinc}(kr)},
\label{kernel}
\end{equation}
where \(r = |\V r|\). This indicates that the TR focused field is isotropic when the receiving antenna is modeled as a point source. Building on \eqref{eq:electric current_field}, we further derive the TR field for a Hertzian dipole \(\tilde{K}({r},  \vartheta)\)
\begin{equation}
\tilde{K}({r},  \vartheta) = \frac{\sin(kr)}{kr}\sin^2\vartheta
	-\frac{\T{j}_1(kr)}{kr}(3\sin^2\vartheta-2),
\label{kernel_vector}
\end{equation}
where $\T{j}_1(kr)$ denotes the first-order spherical Bessel function of the first kind, given by $\T{j}_1(\xi)=\sin(\xi)/\xi^2-\cos (\xi)/\xi$, $\vartheta$ is measured from the dipole polarization direction $\UV e=\UV z$ as in this first example. This shows that the TR focused field is anisotropic when the receiving antenna is modeled as a Hertzian dipole. 

For \eqref{kernel}, the focal spot corresponding to a receiving point source exhibits a 3-dB focal width of $0.44\lambda$, and the first null is located exactly at a displacement of $\lambda/2$ from the focal point.
For \eqref{kernel_vector}, the focal spot corresponding to a receiving Hertzian dipole source shows 3-dB focal widths of $0.578\lambda$ and $0.402\lambda$ at $\vartheta=0^\circ$ and $\vartheta=90^\circ$, respectively, with the corresponding first nulls located at displacements of $0.715\lambda$ and $0.437\lambda$ from the peak position.

The TR approach was originally developed for acoustic systems and is traditionally formulated for scalar fields as in \eqref{kernel}. It is also applicable to electromagnetic radiation problems and can be naturally extended to the full vector-field setting as in \eqref{kernel_vector}, and it corresponds to the imaginary part of the dyadic Green’s function \cite{de2010theory}. However, the CP method exhibits the same behavior for any \(\vartheta\). By contrast, the discrepancy observed along the \(z\)-axis cut mainly arises from the markedly different axial amplitude distributions produced by the two methods.
The comparison of the 2D cuts shows that the optimization results closely match the analytical solutions for both $\vartheta=0^\circ$ and $\vartheta=90^\circ$ planes, and their 2D electric field distribution has a similar profile. For a $z$-polarized receiving Hertzian dipole, the TR-focused spot is broader along the $z$ direction than along the $y$ direction in terms of the 3-dB focal width for \(\vartheta = 90^\circ\).

\begin{figure}[t]
  \centering
    \centering    \includegraphics[width=0.9\linewidth]{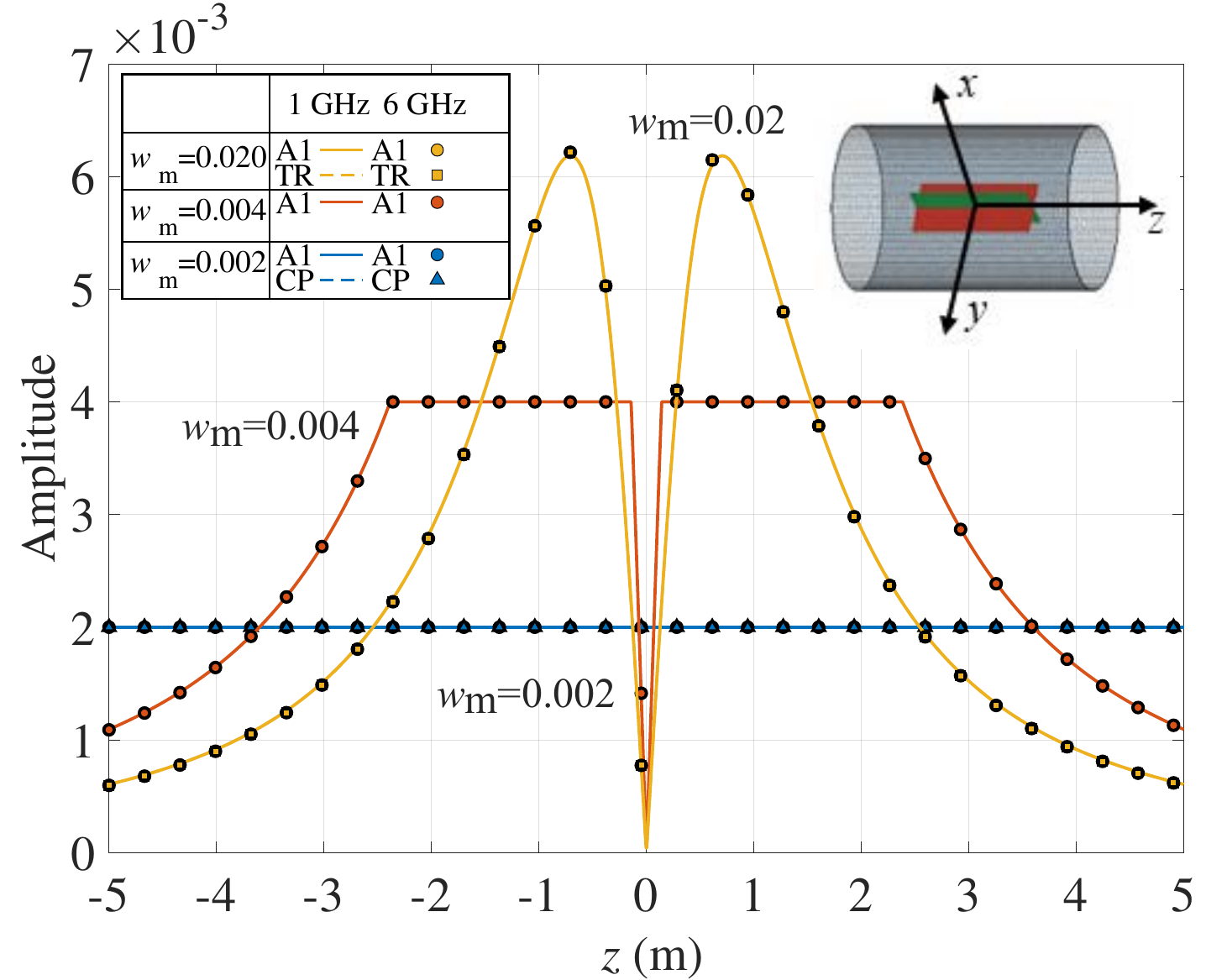}
  \caption{Amplitude distribution of $\M w$ for a cylindrical surface
of radius 1 m and length 10 \(\T{m}\) for \(\phi  = 0^\circ\) when \(\UV e = \hat{\V x}\). \(P_0 = 1~\T W\). The focal point optimized lies at $\V r_\T{f} = \V 0$ with \(x\)-polarization.}
\label{fig:CVX_method_Jz_x}
\end{figure}

Figure \ref{fig:CVX_method_Jz_x}
shows the current amplitude distributions for various \(w_{\T m}\) values when \(\UV e= \hat{\V x}\). Under the global power constraint, the optimized amplitude distribution becomes bimodal along the \(z\)-direction, exhibiting two local maxima. This behavior arises because currents in the immediate vicinity of the focal point contribute negligibly to the \(x\)-polarized field, as the aperture location moves away along \(z\)-axis from the origin, the \(x\)-polarized component increases, but geometric spreading with distance imposes attenuation, yielding a peak followed by a gradual decay. Due to symmetry, an analogous trend holds for the case targeting \(y\)-polarization and remains essentially frequency-invariant. Under both global and local power constraints, the amplitude profile can also develop two flat-top regions.

\begin{figure}[t]
  \centering  \includegraphics[width=1\linewidth]{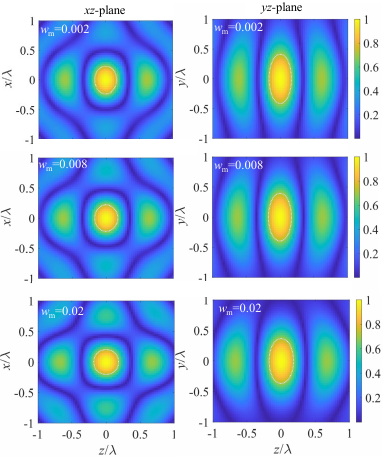}
  \caption{Normalized co-polarized electric field \(E_{\T x}\) optimization result for a cylindrical surface of radius 1 \(\T{m}\) and length 10 \(\T{m}\). The focal point lies at \(\V r_\T{f}=\V 0\) with \(x\)-polarization. The white ring marks the 3 dB boundary of the focal region.
}  \label{fig:global_opt_x_polarization}
\end{figure}

Figure~\ref{fig:global_opt_x_polarization}
presents the \(x\)-polarized electric field distribution at 1\,GHz with the focal point at \(\V r_\T{f} = \V 0\). As the design transitions from the CP method to the TR method, the focal spot is progressively compressed in the $xy$-plane. Under the CP method, the $x$-polarized electric field along the orthogonal $z$-axis cut closely follows the prediction of \eqref{kernel}. 
In contrast, for an $x$-oriented receiving Hertzian dipole, the results of TR co-polarized component agree remarkably well with \eqref{kernel_vector} on both the $xz$ and $yz$-planes for \(\vartheta = 0^\circ\).
 Combined with the previous result for the \(z\)-polarized field, this indicates that different polarizations can produce \(\operatorname{sinc}(\cdot)\)-like gain profiles along their respective orthogonal directions. Compared with the $z$-polarized focusing field, the $x$-polarized field
shows much smaller differences between the two methods. Under the TR method, the
regions with large $x$-directed current amplitude are not those closest to the focal point, so the effective
$x$-directed current distribution over the aperture is less concentrated.
 The $x$-polarized field therefore appears more uniform and closer to that obtained with the \eqref{kernel}.

Additionally, to validate the cross-sectional transformation illustrated in Fig.~\ref{fig:LIS application scenarios indoors}, we simulate a more realistic corridor scenario and compare the electric field intensities obtained for the original rectangular cross-section and for its inscribed and circumscribed circular models. Assuming a rectangular cross section of width \(40\lambda\) and height \(35\lambda\), optimization of \(J_\T z\) alone yields electric field intensities at the focal point \(\V r_\T{f} = \V 0\) of \(0.90\ \mathrm{V/m}\), \(0.98\ \mathrm{V/m}\), and \(1.10\ \mathrm{V/m}\) for the inscribed circle, the rectangle, and the circumscribed circle when the total power is 1 W, respectively. At other sampling points, the field for the rectangle always lies between those of the two circular cases. While complying with the applicable regulatory limits, the electric-field magnitude at the focal point can be further increased by raising the total transmit power.

As shown in \eqref{eq:e-field_expression}, the radiated electric field is given by the superposition of the contributions from electric and magnetic current sources, each weighted by its corresponding dyadic Green’s function. 
Although the two formulations appear formally symmetric in \eqref{eq:e-field_expression}, their associated Green’s-function kernels exhibit different vector structures and polarization couplings in the near field. 
As a result, it is unclear whether electric and magnetic current excitations yield the same near-field focusing performance under the same power constraint.
We therefore compare the field intensities produced by electric currents ($J_{\T z}$, $J_{\T \phi}$) and magnetic currents ($M_{\T z}$, $M_{\T \phi}$) under identical constraints. We impose a single local-power constraint on $J_\T z$ and $J_\T \phi$ (or on $M_\T z$ and $M_\T \phi$) to avoid degenerate cases in which one component remains identically zero at certain spatial locations. The comparison is performed via multipoint sampling at various locations inside the cylinder and for three orthogonal polarizations.

\section{Near-Field Focusing with Discrete Hertzian Dipoles}

Dipole and patch antennas are attractive for LIS or RIS implementations due to their ease of fabrication, excitation, and installation \cite{liang2024filtering}. In this section, we analyze near-field focusing under both the CP and TR methods using discrete Hertzian dipole elements.

We follow the design procedure of the previous section, and validate Hertzian dipole discrete arrays with \(\lambda/2\) element spacing. We model the aperture as multiple concentric circular arrays. Each ring has a radius of \(a = 1~\T{m}\), adjacent elements on the same ring are spaced approximately  \(d=\lambda/2\), and adjacent rings are also separated by \(\lambda/2\). We begin with a \(z\)-polarized (longitudinal  polarization in the Fig.~\ref{fig:LIS_scenario}) Hertzian dipole array. Mutual coupling and scattering are neglected, and only the line-of-sight channel is considered. All simulations are carried out at \(1\,\text{GHz}\) (\(\lambda\approx0.3~\T{m}\)) and \(6\,\text{GHz}\) (\(\lambda\approx0.05~\T{m}\)).

The electric-field approximate expressions for different polarization components and the corresponding
excitation vectors are presented in Appendix~\ref{app:proof_E_components}
Both $z$-polarized and $\phi$-polarized Hertzian dipoles are considered. For the evaluation of the $z$-polarized electric field along the $z$-axis, the $\phi$-polarized component is negligible, so only the $z$-polarized Hertzian dipole is included. We subsequently compare the results at 1 GHz and 6 GHz.

\begin{figure}[t]
  \centering
    \centering    \includegraphics[width=0.9\linewidth]{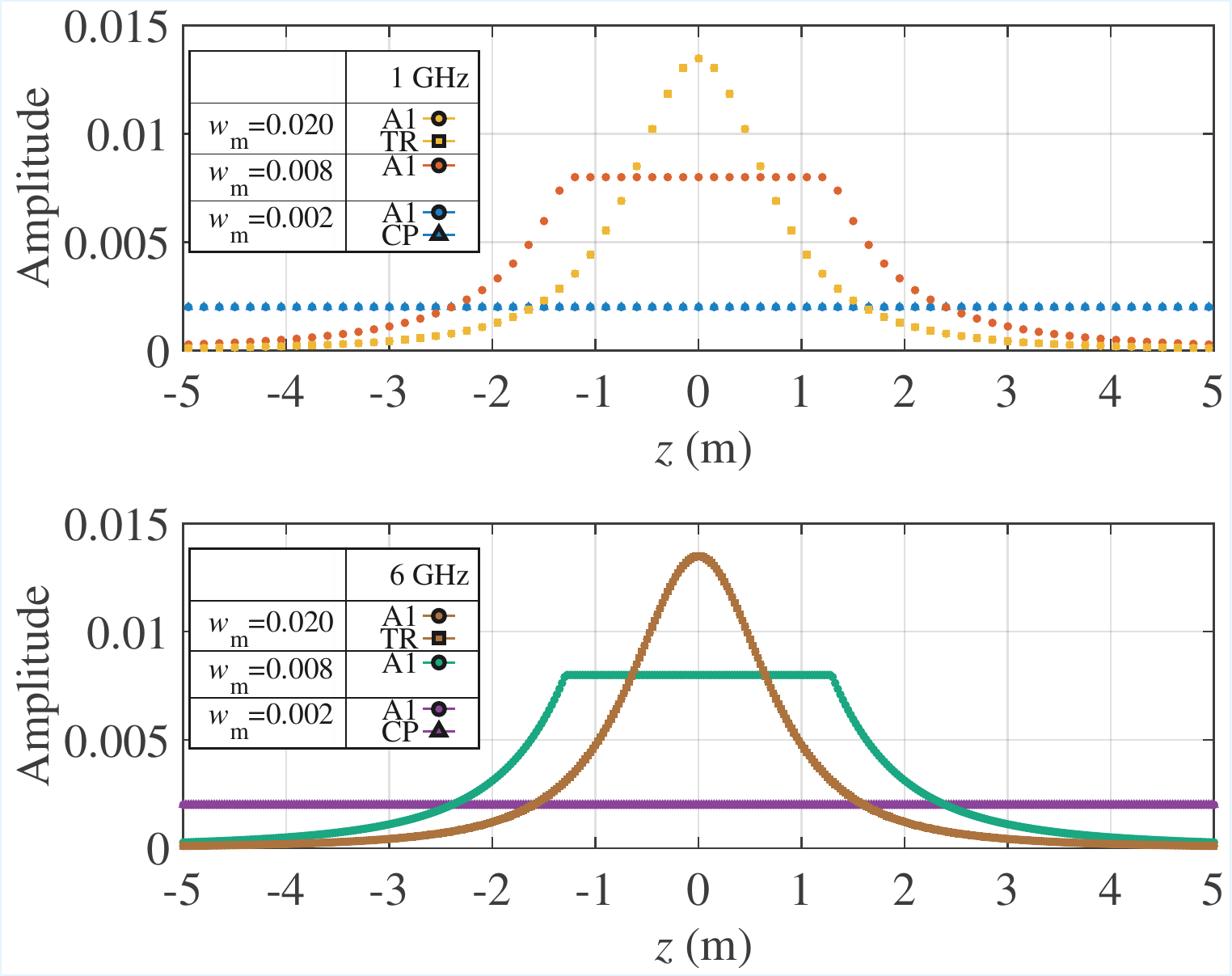}
    \\[-0.5ex]
  \caption{
Discrete distributions of short dipole elements at 1\,GHz and 6\,GHz with \(\lambda/2\) spacing along the \(z\)-direction, the focal point is at $\V r_\T{f} = \V 0$. \(\T{A1}\) denotes Algorithm 1. The $1\,\mathrm{GHz}$ layout 
comprises $42$ rings with $67$ elements per ring, whereas the $6\,\mathrm{GHz}$ layout 
comprises $252$ rings with $401$ elements per ring. Each marker represents an antenna element.
}
\label{fig:global_opt_discrete}
\end{figure}

As shown in Fig. \ref{fig:global_opt_discrete},
for a $10\,\mathrm{m}$ corridor with half-wavelength spacing versus $z$ direction under \(w_{\T m}\!\in\!\{0.02,\,0.008,\,0.002\} A\), 
when enforcing a total-power constraint, we note that the effective surface resistance $R$ 
of the continuous-current model differs from that of a discretized array and also varies 
with element density. Accordingly, the discrete element amplitudes are rescaled by a 
frequency-dependent factor so that the total power matches the continuous solution. 
After this proportional rescaling, the amplitude taper that maximizes the on-axis focal field 
for the discrete arrays aligns with that obtained from the continuous-current. Specifically, in the continuous-aperture model, the surface current density is normalized by enforcing a total transmit-power constraint $\int_S p(w)\,\mathrm{d}S = 2W$, where the local power density is taken proportional to the squared current magnitude, i.e., $p(w)\propto |w|^2$.
 When the cylinder is sampled with
half-wavelength spacing into $N_{\phi}\times M_z$ Hertzian dipoles, the area integral is
approximated by $\int_S |w|^2\,\mathrm{d}S \approx \Delta S
\sum_{\ell=1}^{NM} |w_\ell|^2$, where $\Delta S = S/(NM)$ is the
area of each element. Let
$P_0 \approx \alpha^2 NM / S \propto 1/\lambda^2$, where
$\alpha$ is a normalization constant relating the continuous surface current to
the discrete Hertzian dipole. Thus, the continuous solution can be discretized and used directly as the realizable 
amplitude distribution.

\section{Electric Intensity and Focal Resolution for Extremely Large Array}

In this section, we investigate the near-field focusing behavior of an extremely large-scale array modeled as a discretized Hertzian dipole array, with particular emphasis on two metrics: the focal-spot intensity for different spatial polarizations and the focal resolution. By comparing the performance obtained by the TR and CP approaches, we gain a deeper insight into the practical limitations of LIS in real-world deployments. The analysis focuses on locations along the principal axes of the 3D array, as activating different sub-regions of the LIS, allows the user to be positioned near a principal axis. For example, selecting a segment of the LIS along its length, places the user in a region where the key characteristics remain essentially unchanged. We first analyze the longitudinally polarized array in Fig.~\ref{fig:LIS_scenario}, and then the transversely polarized array.

\subsection{Field Intensity Distribution}
  When the number of elements is large, the summation of the electric fields from discrete array elements at a field point $\V{r}$, as shown in Appendix~\ref{app:proof_E_components} (Eq. (29)) can be accurately replaced by an integral, similar to the continuous-current aperture LIS described in Section~III. This formulation enables efficient evaluation and facilitates the derivation of useful analytical expressions. For each Hertzian dipole element, the amplitude term contains a constant term \(R_{\T {e}} = \ju\eta lk/{4\pi}\) and excitation weight \(w_n\) 
(the excitation current \(I_n\)) is assigned to the element. In the rest of the paper, \(E^{\T{C}}\) denotes the electric field obtained by the CP method, and \(E^{\T{T}}\) represents the electric field obtained via the TR method. The subscripts $x$, $y$, and $z$ on the \(E\) indicate the polarization direction, and the term in parentheses specifies the direction of the focusing axis. When the LIS employs longitudinally polarized antenna elements, the resulting field contains both longitudinal polarization component (corresponding to the co-polarization component \(E_{\T z}\)) and transverse polarization component (corresponding to the cross-polarization components \(E_{\T x}\) or \(E_{\T y}\)). For brevity, we present only the final derived result here, the detailed derivation can be found in Appendix~\ref{app:proof_E_components}.

\begin{figure}[t]
  \centering
    \centering   \includegraphics[width=1\linewidth]{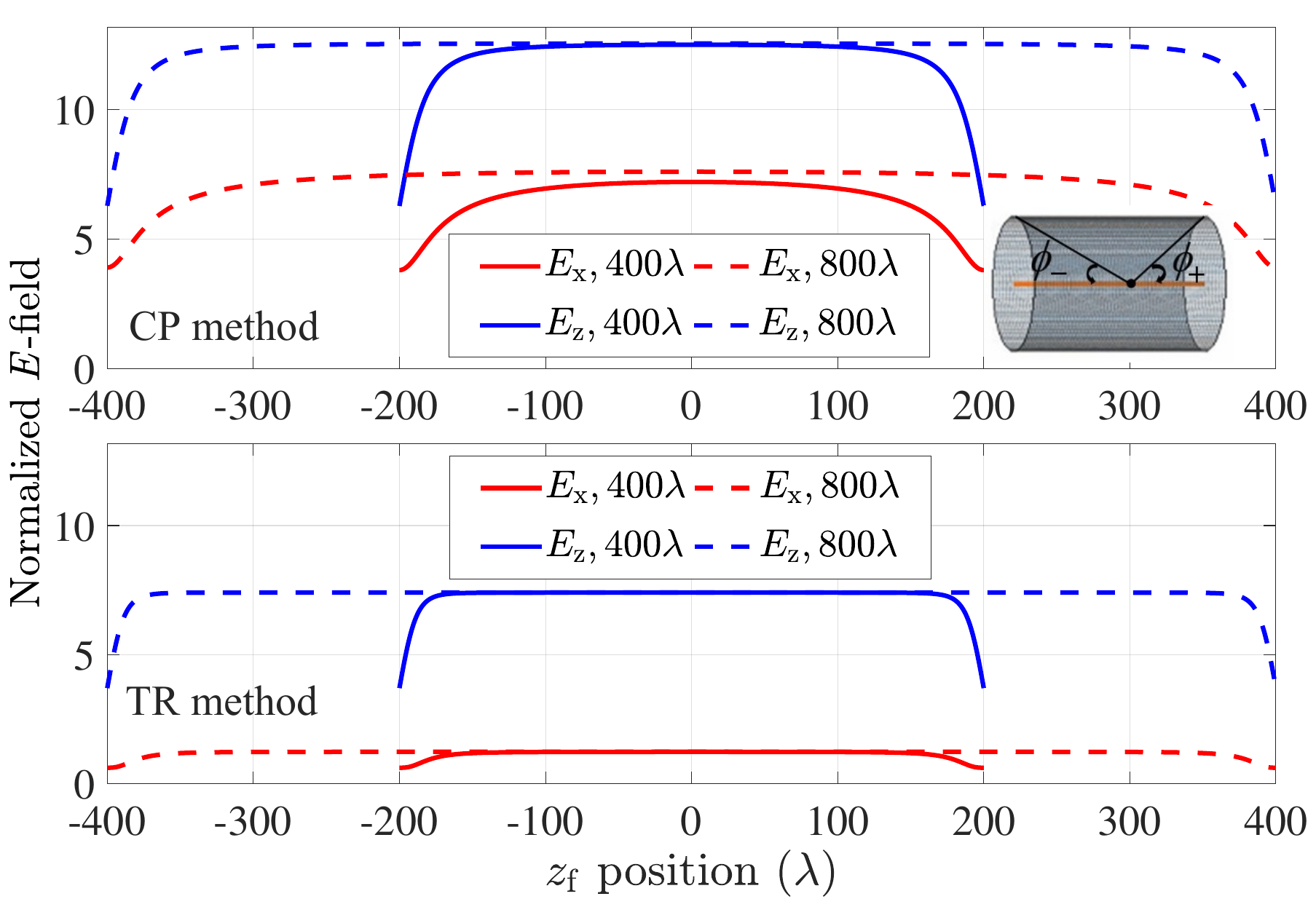}
    \\[-0.5ex]
  \caption{
The normalized peak electric field amplitudes for \( E_{\T x} \) and \( E_{\T z} \) along the \textit{z}-axis.
}
\label{fig:Ex and Ez}
\end{figure}

By applying the CP method \eqref{eq:Ecp} to compensate for the phase terms and setting each current amplitude to \(w_\T{m}\), the normalized focused electric field \(E_{\T z}^{\T C}(z_{\T f})\) is obtained
\begin{equation}
\frac{\lambda^2E_{\T z}^{\T C}(z_{\T f})}{R_{\T {e}} w_\T{m} a}
=\frac{\pi}{2}\cos\phi_++\frac{\pi}{2}\cos\phi_-,
\end{equation}
where $\phi_{+}$ and $\phi_{-}$ are defined as the angles between the $z$-axis and the line segments connecting an arbitrary point on the $z$-axis to the cylinder’s right and left edges, respectively (see Fig.~\ref{fig:Decomposition}) and
a schematic illustration is provided in Fig.~\ref{fig:Ex and Ez}. In this geometry, $\phi_{\pm}\in(0,\pi/2)$. where $\tan\phi_{\pm}={a/\left({L/2\pm z_\T f}\right)}$ (\(L\) is the length along the \(z\) direction, \(a\) is the radius of the cylinder and $z_\T f$ is the position of the focal point at $z$-axis).
This implies that for an extremely large-scale LIS, when $L \gg a$, the expression becomes essentially independent of the position along the $z$-axis and approaches the constant $\pi R_{\mathrm e} w_{\mathrm m} a/\lambda^2$, corresponding to the limiting case $\phi_{\pm}\to 0$.

However, when solving the problem using the TR method, the excitation coefficient \(w_n\) of each antenna element is effectively proportional to its corresponding field amplitude. For the \(E_{\T x}^{\T T}\) component, this implies that \(w_n\) is directly related to the corresponding amplitude attenuation. Based on this principle, in an analogous manner we introduce a scalar factor \(D_{\T r}\), defined as \(D_{\T r} = \sqrt{{2P_0}/({\M h^{\T H}\M R^{-1}\M h})}\) (starting from the global power constraint
\(\M w^{\T H}\M R\,\M w \le 2P_0\),
the optimal solution can be written as \(\M w = D_{\T r}\,\M R^{-1}\M h^{*}\) from \eqref{eq:tr_weight and e-field}
), so we can get
\begin{equation}
\begin{cases}
E_{\T z}^{\T T}(z_{\T f})
= F_2\!\left(\tfrac{L}{2}-z_{\T f}\right)
  - F_2\!\left(-\tfrac{L}{2}-z_{\T f}\right),\\[4pt]
\displaystyle
\frac{\lambda^2 F_2(u)}{D_{\T r} R_{\T {e}} }
= \frac{u a \pi\,(5a^{2}+3u^{2})}{16(a^{2}+u^{2})^{2}}
  + \frac{3\pi}{16}\tan^{-1}\!\left(\frac{u}{a}\right).
\end{cases}
\end{equation}
Similarly, when \(L \gg a\), the result simplifies to \(3\pi^{2}{D_{\T r} R_{\T {e}} /(16\lambda^2)}\). Unlike the expression obtained using the CP method, this result does not depend on \(a\) and \(L\) (except for edge effects near the ends). 

On the other hand, the CP method yields different results for \(E_\T x^{\T C}\) and \(E_\T x^{\T T}\) due to their different formulations.
The resulting expression is given by    
\begin{equation}\label{eq:ExT}
\frac{\lambda^2E_{\T x}^{C}(z_{\T f})}{R_{\T {e}} w_\T{m} a} = 2-\sin\phi_{+}-\sin\phi_{-}.
\end{equation}
For the TR method, it is given as
\begin{equation}
\begin{cases}
E_{\T x}^{\T T}(z_{\T f})
= F_1\!\left(\tfrac{L}{2}-z_{\T f}\right)
  - F_1\!\left(-\tfrac{L}{2}-z_{\T f}\right),\\[4pt]
\displaystyle
\frac{\lambda^2 F_1(u)}{D_{\T r} R_{\T {e}} }
= \frac{\pi a u(u^{2}-a^{2})}{32(a^{2}+u^{2})^{2}}
  + \frac{\pi}{32}\tan^{-1}\!\left(\frac{u}{a}\right).
\end{cases}
\end{equation}
The values of \( E_{\T x}^{\T C} \) and \( E_{\T x}^{\T T} \) converge to \(2 R_{\T {e}}  w_\T{m} a/\lambda^2\) and \(\pi^{2}{D_{\T r} R_{\T {e}} /(32\lambda^2)}\) respectively, as the \(L\) is much larger than \(a\). As before, similar result is obtained for $E_{\T y}^{\T T}$. Figure~\ref{fig:Ex and Ez}
shows the variation of \(E_{\T x}\) and \(E_{\T z}\) with different lengths for the two methods. The parameters \({w_{\T {m}} R_{\T {e}} a /(4\lambda^2)}\) and \({D_{\T r} R_{\T {e}}/(4\lambda^2)}\) are both normalized to~1 (as for Figs. 10, 11 and 12). The orange line inside the cylinder illustrates the corresponding plotting location.
  The ratio of the longitudinal to transverse components remains constant at \(\pi/2\) for the CP method and 6 for the TR method. Furthermore, the corresponding power ratios $\pi/2$ for the CP method and 36 for the TR method serve as key indicators of the polarization energy distribution in these two methods.

\begin{figure}[t]
    \centering
\includegraphics[width=0.97\linewidth]{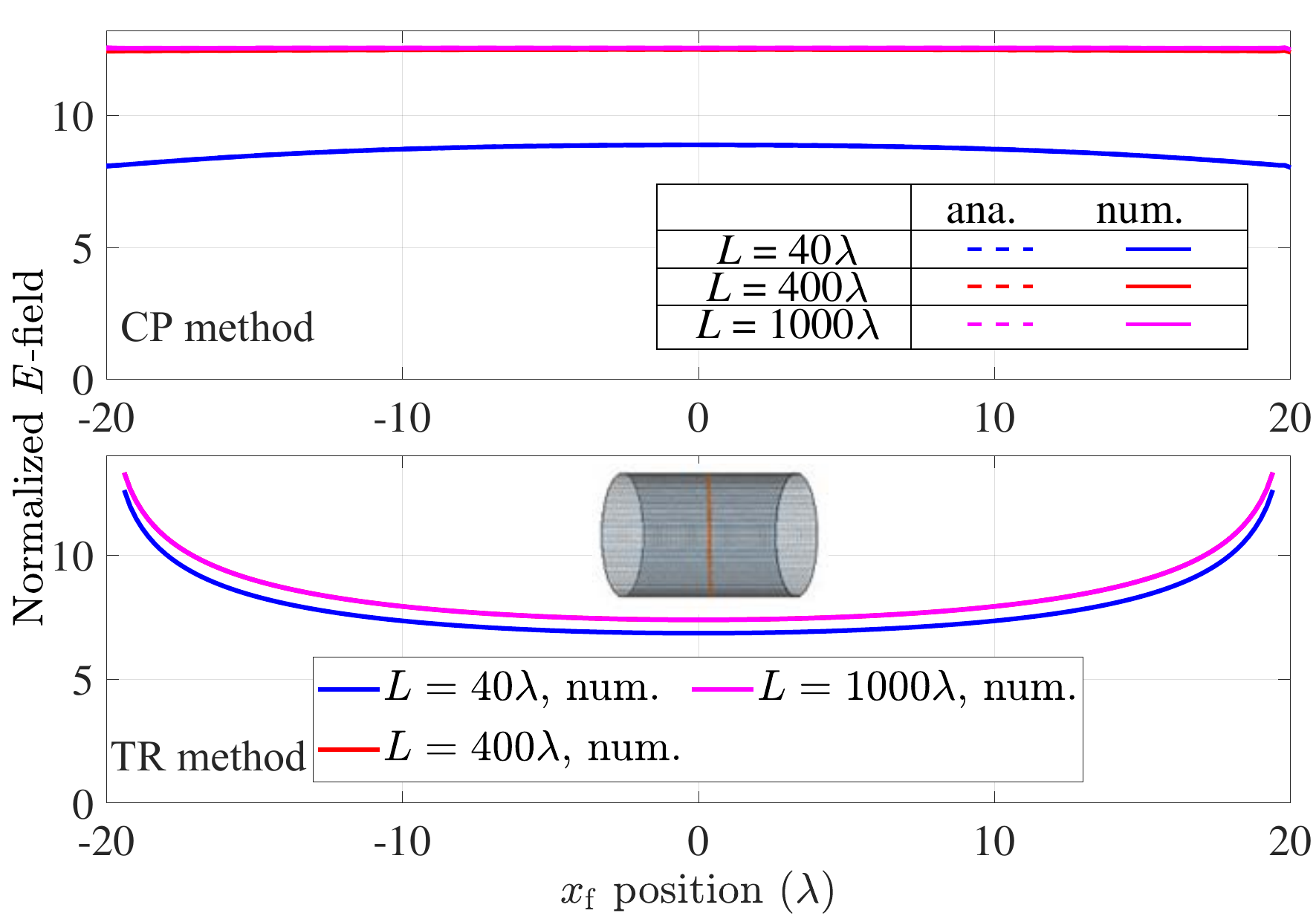}
    \caption{The distribution of the \(E_{\T z}\) component along the central \(x\)-axis. Top, CP method, Bottom, TR method.}
    \label{fig:Ez_CP_TR}
\end{figure}

 Due to the asymmetry of $E_{\T x}$ in the $x$- and $y$-directions, the transverse electric-field
distribution is examined in two cases, with the focal points located along the $x$-axis and
$y$-axis, respectively. The trend of $E_{\T z}$ when the focal point loacated at \( x \)-axis can be simplified to
\begin{equation}
\begin{cases}
\displaystyle
\frac{\lambda^2 E_{\T z}^{\T C}(x_{\T f})}{R_{\T {e}} w_\T{m} a}
= \frac{L}{2}\!\left[
\frac{K_1\!\left(-\dfrac{16ax_{\T f}}{\Delta_{-}}\right)}{\sqrt{\Delta_{-}}}
+
\frac{K_1\!\left(\dfrac{16ax_{\T f}}{\Delta_{+}}\right)}{\sqrt{\Delta_{+}}}
\right],\\[4pt]
\Delta_{\pm} = L^2 + 4(x_{\T f} \pm a)^2.
\end{cases}
\end{equation}
where $K_1(\cdot)$ is the elliptic integral of the first kind \cite{jeffrey2008}.
Figure~\ref{fig:Ez_CP_TR}
compares the analytical and numerical solutions of $E_{\T z}^{\T C}$, as well as the numerical solution of $E_{\T z}$ for the TR method. In contrast to $E_{\T x}$, the $E_{\T z}$ component converges rapidly to a uniform focal level even with a relatively short array. This rapid convergence makes $E_{\T z}$ more practical, since it requires fewer antenna elements to achieve uniform focusing. The TR method converges much faster, even a relatively short array length is sufficient to reach the peak field level within the focal region, which indicates that the TR approach offers greater potential for practical applications. It is worth noting that the TR method becomes invalid when the focal point is placed very close to the cylindrical surface. This is because the elements near the focus operate in its extreme near-field region, leading to abnormally high energy concentrations.

\begin{figure}[t]
    \centering  \includegraphics[width=1\linewidth]{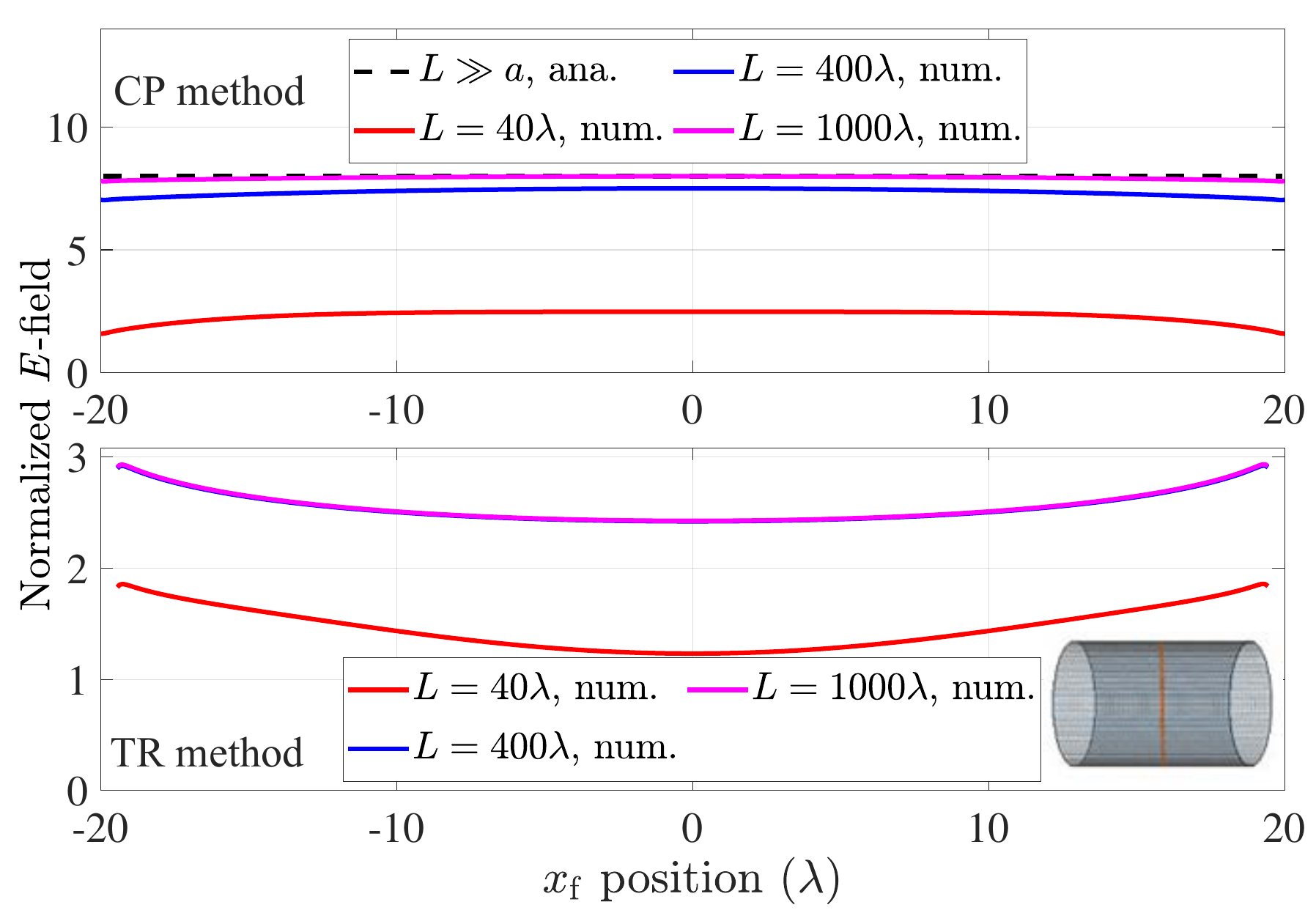} 
    \caption{The distribution of the \(E_{\T x}\) component along the central \(x\)-axis. Ana. means analytical solutions and num. means numerical solutions.}
    \label{fig:Ex_CP_TR}
\end{figure}

In contrast, $E_{\T z}$ is symmetric with respect to $x$ and $y$ directions;
therefore, it is sufficient to consider only the focal point along the $x$-axis. The gain variation of $E_{\T x}$ can be approximately expressed as ($L \gg a$)
\begin{equation}
\frac{\lambda^2 E_{\T x}^{\T C}(x_\T{f})}{R_{\T {e}} w_\T{m} a} \approx
2,
\end{equation}
The numerical solutions for the CP and TR methods, along with the analytical solutions for the CP method are shown in Fig.~\ref{fig:Ex_CP_TR}.
When $L \gg a$, the analytical and numerical results for the CP method agree very well.
 
In TR, once the aperture is large enough, the peak curves for different lengths converge and almost overlap. TR concentrates most power around the focus, and contributions from far elements decay quickly, so enlarging the aperture further brings only small gains. Also, placing the focus closer to the cylindrical surface increases local power concentration and raises the focal field. This is why TR saturates faster than CP, showing that TR mainly benefits from the most effective part of the aperture.

Similarly, by applying the same method, the simplified expression for $E_{\T x}^{\T C}(y_{\T f})$ is obtained
\begin{multline}
\frac{\lambda^2E_{\T x}^{\T C}(y_{\T f})}{R_{\T {e}} w_\T{m} a} = \frac{1}{2y_{\T f}} \Big( 2a + 2y_{\T f} - 2|a - y_{\T f}|  \\
 + \sqrt{L^2 + 4(y_{\T f} - a)^2} - \sqrt{L^2 + 4(y_{\T f} + a)^2} \Big). 
\end{multline}

\begin{figure}[t]
    \centering    \includegraphics[width=1\linewidth]{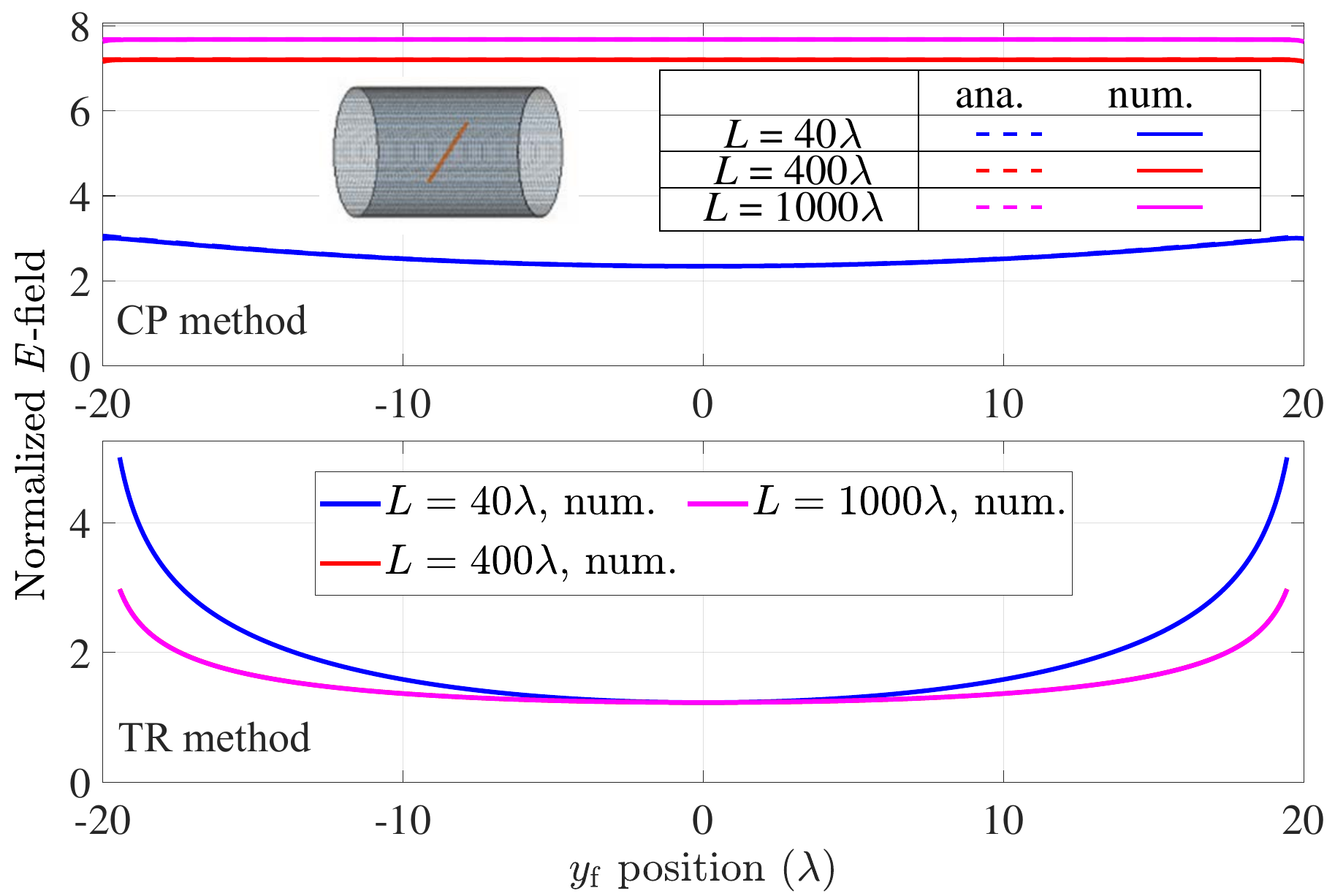} 
    \caption{The distribution of the \(E_{\T x}\) component along the central \(y\)-axis. Top, CP method, Bottom, TR method.}
    \label{fig:Ey_CP_TR}
\end{figure}

Figure~\ref{fig:Ey_CP_TR}
presents the relationship between the position of the focal point along the $\it{y}$-axis and the electric field intensity. The trends between the analytical solution and the numerical solution exhibit a high degree of consistency for arrays of any length. Moreover, when the array size is long enough, the peak focused electric field strength along the $\it{y}$-axis converges to the same result as that along the $\it{x}$-axis. From the analysis in both orientations, once the array length exceeds a critical threshold corresponding in practice to a sufficiently long corridor or room, the maximum focused electric field becomes effectively uniform. Consequently, as a user moves within the enclosed region and the focal point is updated to match the user’s position, the user experiences consistently uniform charging performance or SNR.

\subsection{Beamforming Resolutions for Closed LIS Structure}
Because the TR solution in \eqref{kernel} and \eqref{kernel_vector} is already known, our motivation is to derive the CP formulations. The amplitude of \(E_{\T z}^{\T C}\) (corresponding to the co-polarization component radiated by a $z$-directed Hertzian dipole) at the focal point on the \(z\)-axis varies with longitudinal and transverse displacements according to the following expressions (see Appendix~\ref{app:beam resolution})
\begin{equation}\label{eq:Ez_depth}
\frac{\lambda^2 E_{\T z}^{\T C}(\delta_{\T L})}{R_{\T {e}} w_{\T {m}} a} \approx \\
\frac{\pi}{\sqrt{1+4a^{2}/L^2}}
\operatorname{sinc}\!\left(\frac{\delta_{\T L} k}{\sqrt{1+4a^{2}/L^2}}\right)
,
\end{equation}
and when $L \gg a$,
\begin{equation}
  \frac{\lambda^2 E_{\T z}^{\T C}(\delta_{\T T})}{R_{\T {e}} w_{\T m} a}
  \approx
  \pi \operatorname{sinc}(k \delta_{\T T}),
  \label{Ez_width}
\end{equation}
where $\delta_{\T L}$ and $\delta_{\T T}$ denote the displacements along the longitudinal and transverse directions, respectively, as defined in Fig.~\ref{fig:LIS_scenario}. Accordingly, the $r$ in \eqref{kernel} is the absolute value of $\delta_{\T L}$ and $\delta_{\T T}$ along the corresponding 1D longitudinal and transverse cuts.

As for the \(E_{\T x}^{\T C}\) (corresponding to the cross-polarization component radiated by a $z$-directed Hertzian dipole), the focal depth and the widths along the \(x\) and \(y\) axes are given when $L \gg a$, respectively, by
\begin{equation}\label{Ex_depth}
\frac{\lambda^2 E_{\T x}^{\T C}(\delta_{\T L})}{R_{\T {e}} w_{\T {m}} a} \approx 2 \T H_{-1}(k\delta_{\T L} ) ,
\end{equation}
\begin{equation}\label{eq:Exx_width}
\frac{\lambda^2 E_\T{x}^{\T C}(\delta_{\T T}^{\T x})}{R_{\T {e}} w_{\T {m}} a} \approx
\dfrac{\pi  \T H_0(k\delta_{\T T}^{\T x} )}{ k\delta_{\T T}^{\T x} },
\end{equation}
\begin{equation}\label{eq:Exy_width}
\frac{\lambda^2 E_\T{x}^{\T C}(\delta_{\T T}^{\T y})}{R_{\T {e}} w_{\T {m}} a} \approx
 \frac{2\T {Si} (k \delta_{\T T}^{\T y})}{ k\delta_{\T T}^{\T y} }.
\end{equation}
where \(\delta_{\T T}^{\T x}\) and \(\delta_{\T T}^{\T y}\) represents displacements along transverse directions ($x$- and $y$-axis). $\T H_{-1}(\cdot)$, $\T H_0(\cdot)$ and $\T {Si}(\cdot)$ are the -1-order, 0-order Struve function \cite{babusci2014spherical} and the integral function of $\T {sinc}(\cdot)$. For \eqref{eq:Exx_width} and \eqref{eq:Exy_width}, setting $\delta_{\T T}=0$, then both \({\lambda^2 E_\T{xx}^{\T C}(\delta_{\T T})}/{R_{\T {e}} w_{\T {m}} a}\) and \({\lambda^2 E_\T{xy}^{\T C}(\delta_{\T T})}/{R_{\T {e}} w_{\T {m}} a}\) are equal to 2.
 The Struve functions are closely related with Bessel functions. They solve the corresponding inhomogeneous Bessel equation, and for half–integer orders, they reduce to spherical Bessel functions up to a simple scaling factor \cite{babusci2014spherical}. The expressions \eqref{eq:Ez_depth}--\eqref{eq:Exy_width} are also applicable to conventional corridor scenarios with rectangular cross sections.

\begin{figure}[t]
    \centering
    \includegraphics[width=1\linewidth]{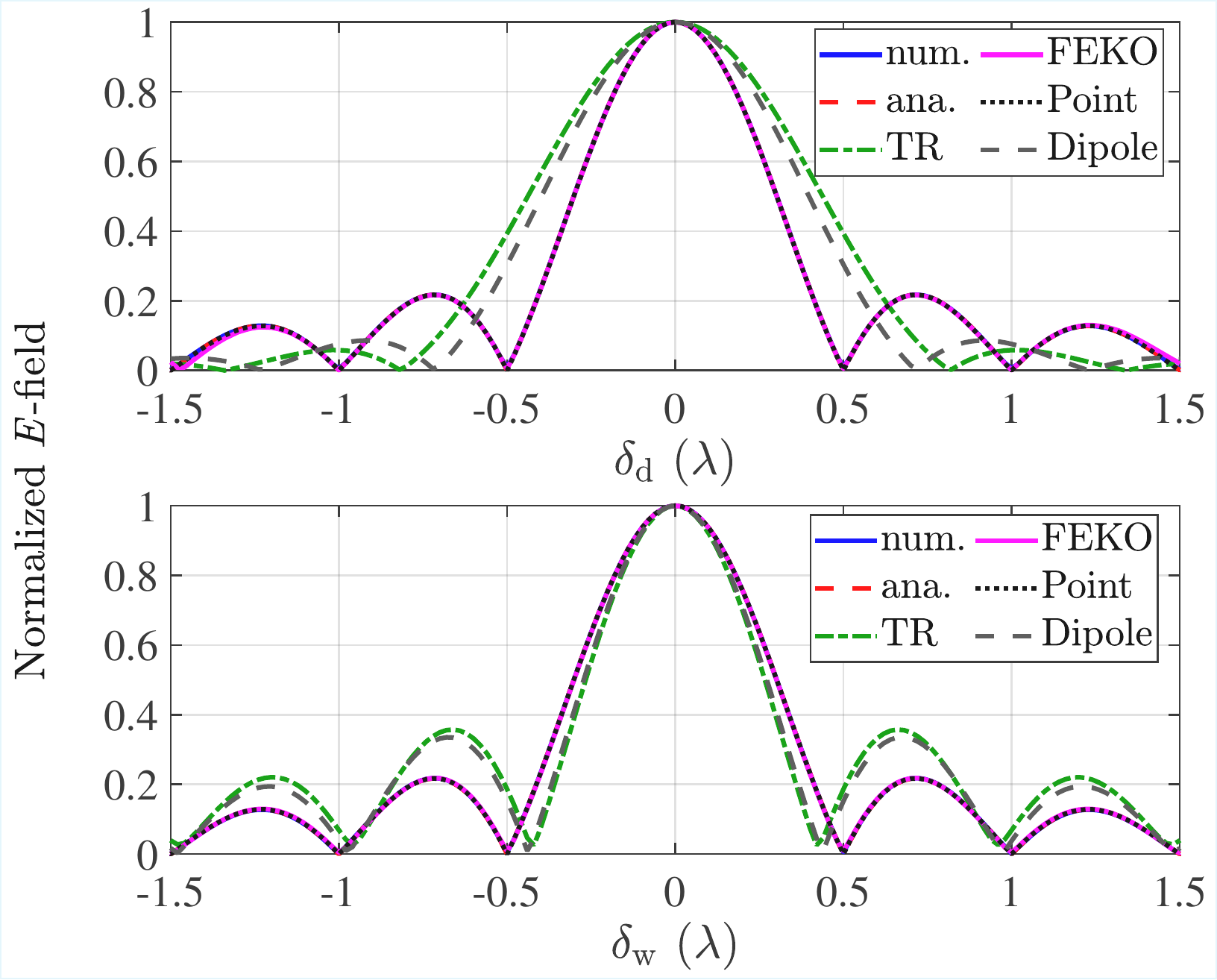} 
    \caption{Normalized electric field amplitude of $E_{\T z}$ versus $\delta_\T{L}$ and $\delta_\T{T}$ for $L = 1000\lambda$. 
Top: variation of $E_{\T z}$ with $\delta_\T{L}$.
Bottom: variation of $E_{\T z}$ with $\delta_\T{T}$. Peak amplitude normalized to 1. ``Point'' follows \eqref{kernel}, while ``Dipole'' follows \eqref{kernel_vector}.
 Unless otherwise specified, the CP method is used by default if TR is not marked in the figure.}
    \label{fig:Ez_resolution}
\end{figure}

\begin{figure}[t]
    \centering    \includegraphics[width=0.93\linewidth]{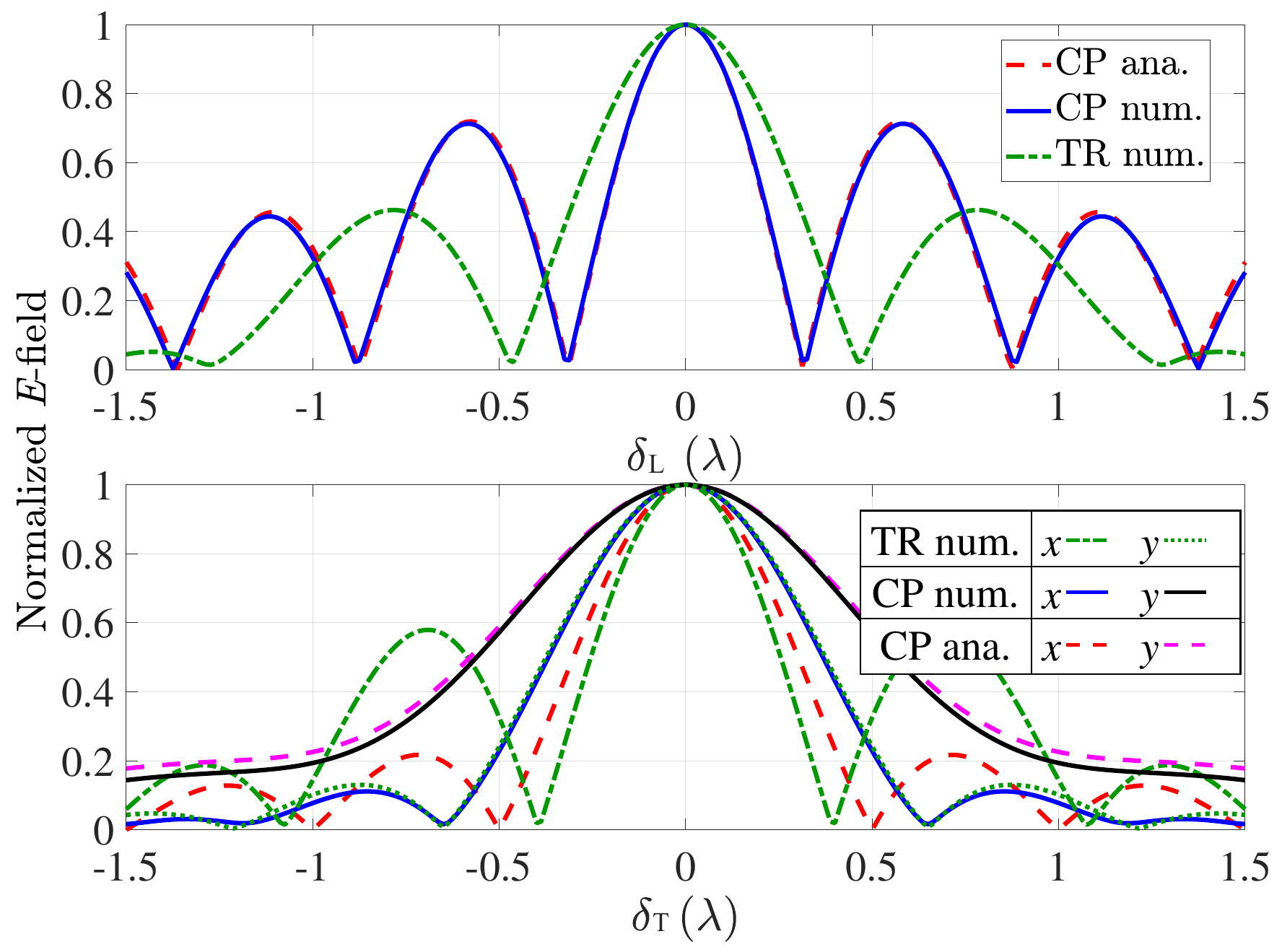} 
    \caption{Normalized electric field amplitude of $E_\T{x}$ versus $\delta_\T{L}$ and $\delta_\T{T}$ for $L = 1000\lambda$. 
Top: variation of $E_\T{x}$ with $\delta_\T{L}$.
Bottom: variation of $E_\T{x}$ with $\delta_\T{T}$. Peak amplitude normalized to 1.}
    \label{fig:Ex_resolution}
\end{figure}

As shown in Figs.~\ref{fig:Ez_resolution}
and~\ref{fig:Ex_resolution},
the numerical solution matches the analytical solution \eqref{eq:Ez_depth}--\eqref{eq:Exy_width} very well for the focal resolution. Figure~\ref{fig:Ez_resolution} also presents the results obtained from 2024 Altair FEKO software, which employed 401 (10 \(\T{m}\) length corresponding to 6 GHz) rings with a radius of 1 \(\T{m}\). A high level of consistency is observed in the results. In addition, the results of \eqref{kernel}--\eqref{kernel_vector} are normalized to unit amplitude, where the variable $r$ in \eqref{kernel}--\eqref{kernel_vector} corresponds to $\delta_\T T$ in the present notation. 
The 3-dB focal width and the focal depth of the \(E_{\T z}\) focal point at the focus are nearly equal, both being \(0.44\lambda\), which indicates that the focal shape appears circular. Meanwhile, this also validates the electric field distribution shown in Figs.~\ref{fig:z_polarzation_16} and~\ref{fig:global_opt_x_polarization}. It is worth noting that the 3 dB focal width produced by TR method is narrower than that of the CP method, from an energy perspective, the TR excitation allocates a larger fraction of the radiated power into the vicinity of the focal region, while reducing the power spread in the surrounding area. From the closed TR cavity result in \eqref{kernel}, we observe that the CP-based result of the present design is very similar. In the numerical results, some higher-order terms in $r$ were neglected in (12), which leads to small discrepancies in the sidelobes, whereas the main lobe agrees well with the analytical solution. Moreover, the CP result in \eqref{Ez_width} has the same functional form, which further confirms that the theoretical focal point achieves the optimal transverse resolution.

Unlike the trend observed in \(E_{\T z}\), \(E_{\T x}\) exhibits more high-level sidelobes along the \(z\)-axis direction, meaning that multiple focal regions appear along the focal longitudinal resolution. The focus transversal resolution of the main lobe is \(0.31\lambda\), whereas along the \(x\)- and \(y\)-axis directions, the focus expands with focus transversal resolutions of \(0.54\lambda\) and \(0.84\lambda\), respectively.

 In Fig.~\ref{fig:Ez_resolution}, the TR method exhibits a peak gain trend similar to that of the CP method, namely that the electric field remains constant along the cylindrical axis except near the edges. When the focal point is located at other positions along the $z$-axis (except near the edges), the results remain essentially consistent. The 3-dB focal longitudinal resolution is approximately $0.66\lambda$, while the 3-dB focal transversal resolution is approximately $0.4\lambda$, which is quite close to the results from \eqref{kernel_vector}. It is worth noting that this result corresponds to the converged outcome for near-field focusing with extremely large arrays.
From this perspective, what differs markedly is that the resolution obtained with the TR method is no longer symmetric in the two dimensions. In particular, the 3-dB focal longitudinal resolution deteriorates, while the focus width resolution improves. However, this improvement comes at the cost of an approximately twofold increase in sidelobe level, which in turn introduces additional energy loss. However, the foregoing analysis reveals that when the global power constraint dominates, the TR method can achieve a field strength several times greater than that of the CP method, implying a substantial improvement in efficiency. We also analyze the transversely polarized case which is shown in Appendix \ref{app:Transversely Polarized}.

\section{Conclusions}
This work investigated near-field focusing with a fixed-size aperture LIS across frequencies and polarizations under both local and global power constraints. For continuous apertures, the optimal current magnitude distribution matches the TR solution under global constraints and the CP solution under dominant local constraints, and remains frequency-invariant for a fixed-size aperture. The focal intensities of the three polarization components under TR excitation remain nearly uniform inside the cylindrical region, and the corresponding field gains approach constant values, with noticeable deviations only near the edges. In contrast, under CP excitation these values vary with the aperture size.
 We further established one focusing cases where polarized field components maintain fixed ratios and where \(z\)-polarizations yield equal width and depth of focus. These theoretical insights were validated using discretized Hertzian dipole implementations and full-wave FEKO simulations.

 \appendices
\section{
Green’s‐function matrix \(\M{h}\) in rectangle LIS panel}
\label{app:proof_A}
Let the filed point be $\V r=(x,y,z)$. 
The rectangular panel lies on $z=0$ with region 
$S=\{(x',y',0)\mid x'\in[-L_x/2,L_x/2],\; y'\in[-L_y/2,L_y/2]\}$.
Expand the tangential surface current density on a rectangular grid using
basis functions
\begin{equation}
\V J(\V r')
= \sum_{m=1}^{N_x N_y}\!\Big[w_m^x\,\V \psi_m^x(\V r')
\;+\; w_m^y\,\mathbf \psi_m^y(\V r')\Big],
\end{equation}
where $\V \psi_m^{(x)}$ and $\V \psi_m^{(y)}$ are the $x$- and $y$-directed
tangential basis functions supported on the $m$-th cell $S_m$. The scalar field component at $\V r$ along a given polarization
$\UV e$
\begin{equation}
E(\V r_0,\UV e)
= \UV e \cdot \!
\int_{S'} \V{G}_\T{J}^\T{E}(\V{r}, \V{r}')\V J(\V r')\, \diff \T S'.
\end{equation}
Substituting the expansion and collecting coefficients yields
\begin{equation}
E(\V r_0,\UV e)=
\underbrace{\big[A^{x}_{1\!:\!N},\,A^{y}_{1\!:\!N}\big]}_{\M h}\,
\underbrace{\big[w^{x}_{1\!:\!N},\,w^{y}_{1\!:\!N}\big]^{\T T}}_{\M w},
\end{equation}
where $N=N_xN_y$. The coefficient for each cell is
\begin{equation}
h_m^{\alpha} =
\UV e \cdot
\iint_{S_m}
\V{G}_\T{J}^\T{E}(\V{r_0}, \V{r}')
\V \psi_m^{\alpha}(\V r')\, \diff S,
\\ \alpha\in\{x,y\}.
\end{equation}
Similarly, for the electric field produced by a magnetic current, the coefficient for each cell is
\begin{equation}
h_{m,(M)}^{\alpha}
= \UV e \cdot
\iint_{S_m}
\V{G}_\T{M}^\T{E}(\V{r}, \V{r}')
{\V \varphi}^{\alpha}_{m}\!\big(\V r'\big)\, \diff S, \\ \alpha\in\{x,y\}.
\end{equation}
where $\V \varphi^{\alpha}_{m}$ is the basis function for magnetic current.

\section{Algorithm~\ref{alg:proposed}}\label{app:Algorithm 1}
The Algorithm~\ref{alg:proposed} can be seen from the algorithm flow. It is used to find the optimal current density distribution under both local power and total power constraints.
\begin{algorithm}[t]
  \caption{Sequential Global\&Local Algorithm}
  \label{alg:proposed}
  \begin{algorithmic}[1]
    \Require $\M h\!\in\!\mathbb{C}^{3\times L}$ (Green function matrix), $\V e\!\in\!\mathbb{C}^3$ (target polarization),
             $P_0\!>\!0$, $w_{\T m}\!>\!0$,
             $\varepsilon\!>\!0$ (tolerance), $K_{\max}\!\in\!\mathbb{N}$ (max iters)
    \State $\V g \gets \V e^{\mathrm H}\M h \in\mathbb{C}^{1\times L}$ 
    \If{$\|\V g\|_2<\varepsilon$} \Return $\mathbf{0}$ \EndIf
    \State $ \V v \gets {\V g}^{\T H}/\|\V g\|_2 $
    \State Define $\mathrm{clip}(\V v,w_{\T m}) \triangleq \min\{|\V v|, w_{\T m}\}$
    \State Define $\phi(\beta)\triangleq\big\| \mathrm{clip}(\beta \V v,\,w_{\T m}) \big\|_2^2$
    \State \textbf{Bounds init:} $\beta_{\rm lo}\!\gets\!0,\;\beta_{\rm hi}\!\gets\!\sqrt{P_0}/\|\V v\|_2$
    \While{$\phi(\beta_{\rm hi})<P_0$}  \label{line:doubling}
      \State $\beta_{\rm hi}\gets 2\,\beta_{\rm hi}$
    \EndWhile
    \For{$k=1$ \textbf{to} $K_{\max}$}
      \State $\beta\gets(\beta_{\rm lo}+\beta_{\rm hi})/2$
      \If{$\phi(\beta)\ge P_0$} $\beta_{\rm hi}\gets\beta$ \Else \ $\beta_{\rm lo}\gets\beta$ \EndIf
      \If{$\big|\phi(\beta)-P_0\big|\le \varepsilon$} \textbf{break} \EndIf
    \EndFor
    \State \Return $\V w\gets \mathrm{clip}(\beta \V v,\,w_{\T m})$
  \end{algorithmic}
\end{algorithm}

\section{Electric field for different polarizations}\label{app:proof_E_components}
According to the definition of the near/far field boundary \cite{7590187}, the reactive near-field region of a single element is extremely short. Therefore, we retain only the first-order term in $1/{\lvert\V{r}- \V{r}'\rvert}$ in \eqref{eq:electric current_field}.
Summing over all array elements yields the following approximate expression:
 \begin{equation}\label{eq:dipole_approximation}
 	\V{E}(\V{r})=\sum_{i=1}^N\frac{\ju\eta_0 I_n lk}{4 \pi r_n}\eu^{-\ju kr_n}(\UV r_n\times(\UV p\times\UV r_n)),
 \end{equation}
 where $\UV r_n$ is the unit vector from the source point to the field point for the \(n\)-th element.
$I_n$ is the current excitation amplitude of the \(n\)-th element, and \( l \) is the Hertzian dipole’s length.

\begin{figure}[t]
  \centering
  \begin{minipage}{0.8\linewidth}
    \centering
    \includegraphics[width=1\linewidth]{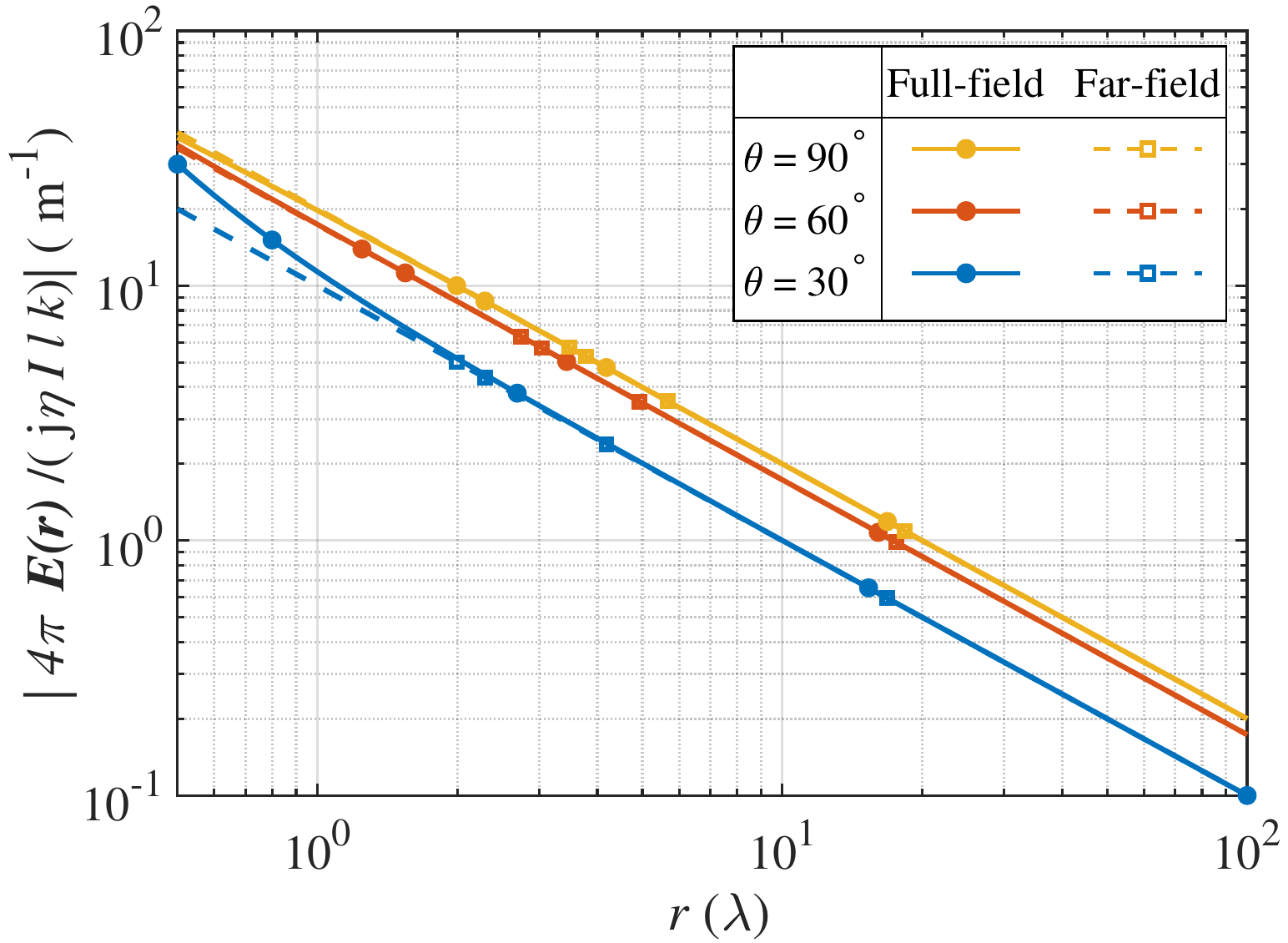}
    \\[-0.5ex]
  \end{minipage}
  \caption{The dyadic Green’s function \eqref{eq:electric current_field} is compared with its far-field approximation \eqref{eq:dipole_approximation}. $\theta$ denotes the angle between $\UV r$ and $\UV p$.}
\label{fig:fullform_approximation}
\end{figure}

Figure \ref{fig:fullform_approximation}
illustrates a comparison between the dyadic Green’s function in \eqref{eq:electric current_field} and its far-field approximation in \eqref{eq:dipole_approximation} under varying $\theta$ (the angle between $\UV r$ and $\UV p$) for an individual element. For observation distances greater than one wavelength, the discrepancy between them is negligible. As a result, the user typically lies in what is conventionally considered 
the far-field region of the individual element. However, when considering an array, the user still fall within 
the near-field region of the LIS panel.   It is worth noting that $\UV e$ denotes the polarization direction at the focal point, whereas $\UV p$ denotes the orientation of the Hertzian dipole moment.

  For the analysis of near-field behavior, using a Cartesian coordinate system enables a more comprehensive characterization of the spatial distribution of focusing properties in three dimensions. Accordingly, the vector identity \(\UV r\times(\UV p\times\UV r) = \UV p - (\UV r\cdot\UV p)\UV r\) states that the cross product of $\UV r$ with $\UV p\times\UV r$ equals $\UV p$ minus the projection of $\UV p$ onto $\UV r$ times $\UV r$.  The position of each element is $\ a_n = \left( a \cos \phi_n, a \sin \phi_n, (m-1)d \right), \quad \phi_n = 2\pi n/{N}$, \( n \) is the index of each element in a circular array, \( m \) is the number of the circular rings, \(m,n\in\mathbb{Z}\). Because the circumference \(2\pi a\) is not an integer multiple of \(\lambda/2\) at either \(1\,\mathrm{GHz}\) or \(6\,\mathrm{GHz}\), the ring is closed with \(N\) uniformly spaced elements, yielding a final inter-element spacing that is slightly below \(\lambda/2\).
Based on \eqref{eq:dipole_approximation}, the expressions for the electric field components for three polarizations can be derived as follows
\begin{equation}\label{eq:H_EX expression}
E_{\T x} = \sum_{m=-M/2}^{M/2} \sum_{n=1}^{N} \left[(m-1)d - z \right] \left( x - a \cos \phi_n \right) Q_{mn}(x, y, z),
\end{equation}
\begin{equation}\label{eq:H_EY expression}
E_{\T y} = \sum_{m=-M/2}^{M/2} \sum_{n=1}^{N} \left[(m-1)d - z \right] \left( y - a \sin \phi_n \right) Q_{mn}(x, y, z),
\end{equation}
\begin{equation}\label{eq:H_Ez expression}
E_{\T z} = \sum_{m=-M/2}^{M/2} \sum_{n=1}^{N} P_n(x, y) Q_{mn}(x, y, z).
\end{equation}
\( Q_{mn}(x, y, z) \) represents a common factor shared by the three formulas, which includes both the amplitude term and the phase term. Their expressions are given as:
\begin{align}
Q_{mn}(x, y, z) &= \frac{R_{\T {e}} I_0\eu^{-\ju k \sqrt{P_n(x, y) + (d_m - z)^2}}}{\big[P_n(x, y) + (d_m - z)^2\big]^{3/2}},\\
P_n(x, y) &= (x - a \cos \phi_n)^2 + (y - a \sin \phi_n)^2, 
\end{align}
where \( R_{\T {e}} = \ju \eta l k / (4 \pi) \) and \( d_m = (m-1)d \).
The expression of the electric field components reveals that each component contains a distinct and unique part. Then for the next step, we need to 
assign the steering vector on \eqref{eq:H_EX expression}, \eqref{eq:H_EY expression}, \eqref{eq:H_Ez expression}, is given by
\begin{equation}
\mathbf a(m,n)
=
\begin{bmatrix}
\eu^{\ju k r_{1,1}} & \eu^{\ju k r_{1,2}} & \cdots & \eu^{\ju k r_{1,n}} \\[6pt]
\eu^{\ju k r_{2,1}} & \eu^{\ju k r_{2,2}} & \cdots & \eu^{\ju k r_{2,n}} \\[3pt]
\vdots          & \vdots          & \ddots & \vdots          \\[3pt]
\eu^{\ju k r_{m,1}} & \eu^{\ju k r_{m,2}} & \cdots & \eu^{\ju k r_{m,n}}
\end{bmatrix},
\end{equation}
where
\begin{equation}
r_{m,n} = \sqrt{(x - a\cos\phi_n)^2 + (y - a\sin\phi_n)^2
           + \bigl(z - d_m\bigr)^2}\,.
\end{equation}
Here, we take $E_{\T x}^{\T C}(z_{\T f})$ as an example to illustrate how its closed-form expression can be obtained. From \eqref{eq:H_EX expression}, we can derive the following when the number of elements is large enough
\begin{equation}\label{eq:zf integration}
\frac{\lambda^2E_{\T x}^{\T C}(z_{\T f})}{R_{\T {e}} w_\T{m}a} =\;
    \frac{a}{4}\int_{-\frac{L}{2}}^{\frac{L}{2}}
    \int_{0}^{2\pi}
    \frac{\left| l - z_{\T f} \right| \left| \cos\phi \right|
    }{
        \lvert\V{r}- \V{r}'\rvert^{3}
    }
    \diff\phi\diff l,
\end{equation} 
where \(\V{r}'\) is employed in the cylindrical coordinate system. According to the definition of the CP method \eqref{eq:Ecp}, the field contributions of all elements are added in phase at the focal point. Therefore, in \eqref{eq:zf integration} the absolute value of each term in the numerator is taken prior to integration, and the same convention is applied when computing the remaining field components. Then we can get the result
\begin{equation}\label{Ex}
\frac{\lambda^2E_{\T x}^{C}(z_{\T f})}{R_{\T {e}} w_\T{m} a} = 2 - 
\frac{1}{\sqrt{1 + \left( \frac{L/2 + z_{\T f}}{a} \right)^2}} 
- 
\frac{1}{\sqrt{1 + \left( \frac{L/2 - z_{\T f}}{a} \right)^2}}.
\end{equation}
It is worth noting that the denominators of the second and third terms in the \eqref{Ex} include factors that depend on the angle between the line connecting the focal point to the cylinder rim and the $z$-axis, then $E_{\T x}^{\T C}(z_{\T f})$ in \eqref{eq:ExT} can be obtained.

\section{Electric field components for longitudinally polarized Hertzian dipole array}\label{app:beam resolution}

For \eqref{eq:Ez_depth}–\eqref{eq:Exy_width}, the procedure for obtaining the focus resolution is analogous. Therefore, we provide a detailed explanation for \eqref{Ez_width}. For \eqref{eq:H_Ez expression}, applying the CP approach yields the following expression for the phase term \(p\)
\begin{equation}
p = \eu^{-\ju k\!\left(
\sqrt{a^{2} + (d_m - z)^{2}}
-\sqrt{\delta_{\T T}^{2} - 2\delta_{\T T} a\cos\theta + a^{2} + (d_m - z)^{2}}
\right)},
\end{equation}

Since $\delta_{\T T}$ is a small quantity, we expand both square-root terms in a Taylor series and retain only the first-order term, which yields
\begin{equation}
p \approx \exp(
-\ju k \delta_{\T T} a \cos\phi/\sqrt{d_m^{2}+a^{2}-2 d_m z+z^{2}}
),
\label{eq:phase_taylor}
\end{equation}
combining this with the amplitude integral, we obtain
\begin{multline}
\frac{4\lambda^2 E_{\T z}^{\T C}(\delta_{\T T})}{ aR_{\T {e}} w_{\T {m}}}
\approx 
\int_{-L/2}^{L/2} \int_{0}^{2\pi}
\bigl(\delta_{\T T}^{2}-2\delta_{\T T} a\cos\phi + a^{2}\bigr)
\\
p\bigl(\delta_{\T T}^{2}-2\delta_{\T T} a\cos\phi + a^{2} + (d_m-z)^{2}\bigr)^{-3/2}
\, \diff \phi\, \diff l.
\end{multline}

Taking into account that $\delta_{\T T}$ is small, the contribution of this term to the overall result is negligible, and the expression can be further simplified as
\begin{equation}
\frac{4\lambda^2 E_{\T z}^{\T C}(\delta_{\T T})}{R_{\T {e}} w_{\T {m}} a}
\approx \int_{-L/2}^{L/2}
\frac{
2\pi a^{2}
J_{0}\!\left(
\frac{\delta_{\T T} k a}{\sqrt{a^{2} + (d_m - z)^{2}}}
\right)
}{
\bigl(a^{2} + (d_m - z)^{2}\bigr)^{3/2}
}
\, \diff l,
\end{equation}
 where $J_0(\cdot)$ denotes the zeroth-order Bessel function of the first kind.
After suitable changes of variables (first to the radial variable
\(u_1=\sqrt{a^{2}+(d_m-z)^{2}}\), and then to the angular variable
\(u_2 = a\sec\phi'\)), the integral can be rewritten in the compact form (as $L \to \infty$
)
\begin{equation}
\frac{4 \lambda^2 E_{\T z}^{\T C}(\delta_{\T T})}{R_{\T {e}} w_{\T {m}} a}
\approx 2\pi \int_{0}^{\pi/2}\cos\phi'\,J_{0}\!\bigl(\delta_{\T T} k\cos\phi'\bigr)\,\diff \phi',
\end{equation}
and the expression \eqref{Ez_width} is then obtained.

\section{Analysis of Transversely Polarized Antenna}\label{app:Transversely Polarized}
\begin{figure}[t]
    \centering    \includegraphics[width=1\linewidth]{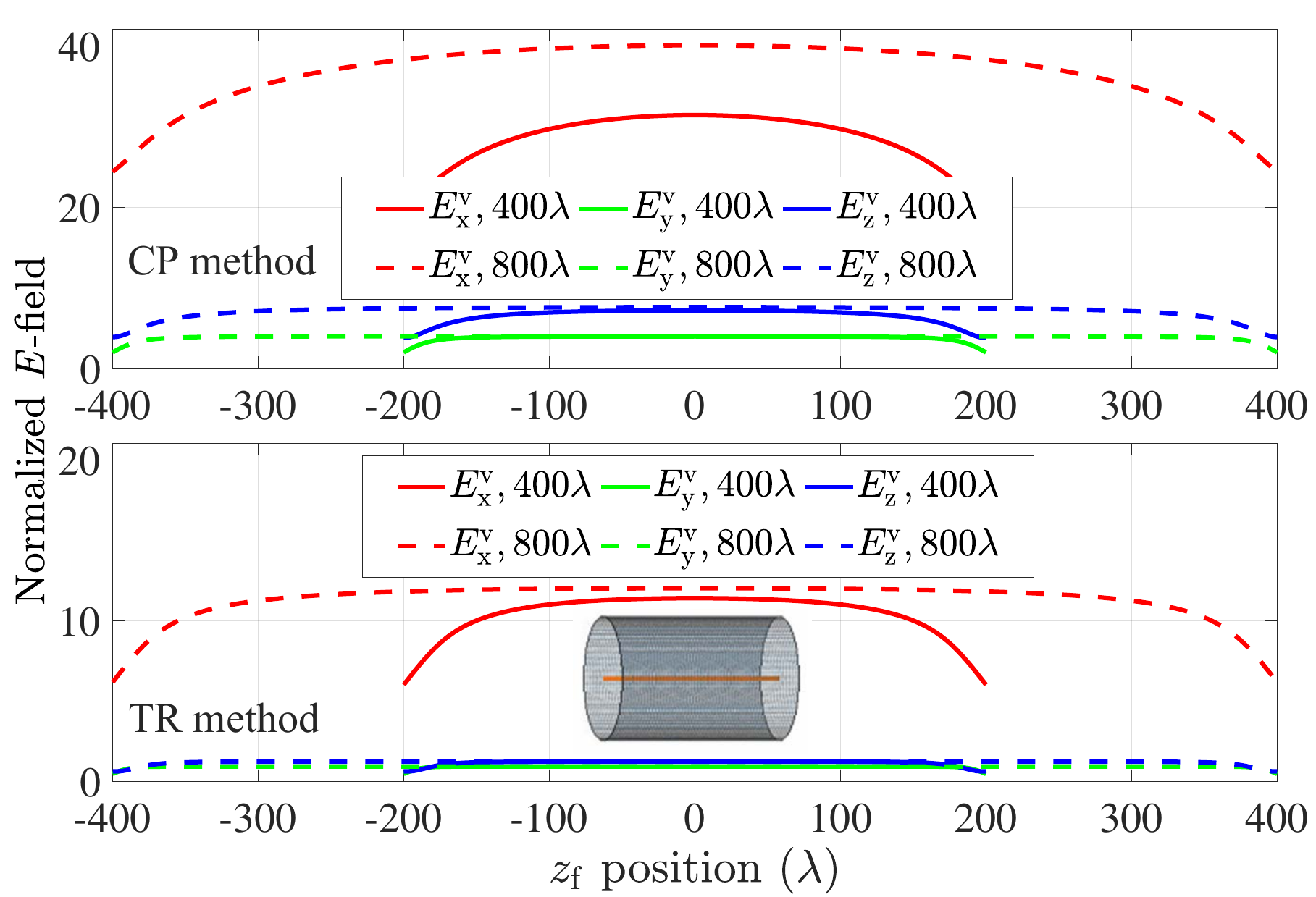} 
    \caption{The normalized peak electric field amplitudes for \( E_{\T x} \), \( E_{\T y} \) and \( E_{\T z} \) along the \textit{z}-axis for transverse polarization with different \(L\). The parameters \({w_{\T {m}} R_{\T {e}} a /(4\lambda^2)}\) and \({D_{\T r} R_{\T {e}} /(4\lambda^2)}\) are both normalized to~1. The numerical results agree well with the analytical solution.
The orange line inside the cylinder illustrates the corresponding plotting location.
}
    \label{fig:vertical gain}
\end{figure}

For the transversely polarized Hertzian dipole antenna, the corresponding different electric field components can be obtained (see Appendix~\ref{app:vertical hertizian dipole}). The closed-form solutions for the peak gain in these three cases are also provided in Appendix~\ref{app:vertical hertizian dipole}, with the corresponding curves shown in Fig.~\ref{fig:vertical gain}.
Interestingly, unlike the longitudinally polarized case, the field distribution for the main polarization is significantly larger than the other two polarizations. Combining \eqref{eq:V_EY integration} and \eqref{eq:V_EZ integration} (see Appendix~\ref{app:vertical hertizian dipole}) shows that the normalized cross-polarized components $E_{\T y}^{\T v}$ and $E_{\T z}^{\T v}$ asymptotically approach \({w_{\T {m}} R_{\T {e}} a /{\lambda^2}}\) and \({2w_{\T {m}} R_{\T {e}} a /{\lambda^2}}\) under the CP method, respectively, whereas the co-polarized component $E_{\T x}^{\T v}$ increases with $L$. Consequently, its peak can far exceed the focusing intensity obtained from longitudinal polarization. Under the TR scheme, the results exhibit a similar trend. The main polarization component is more than one order of magnitude stronger (over ten times larger) than the other two polarizations, while the peak gain is noticeably smoother compared with that obtained using the CP method.

\begin{figure}[t]
    \centering
    \includegraphics[width=0.93\linewidth]{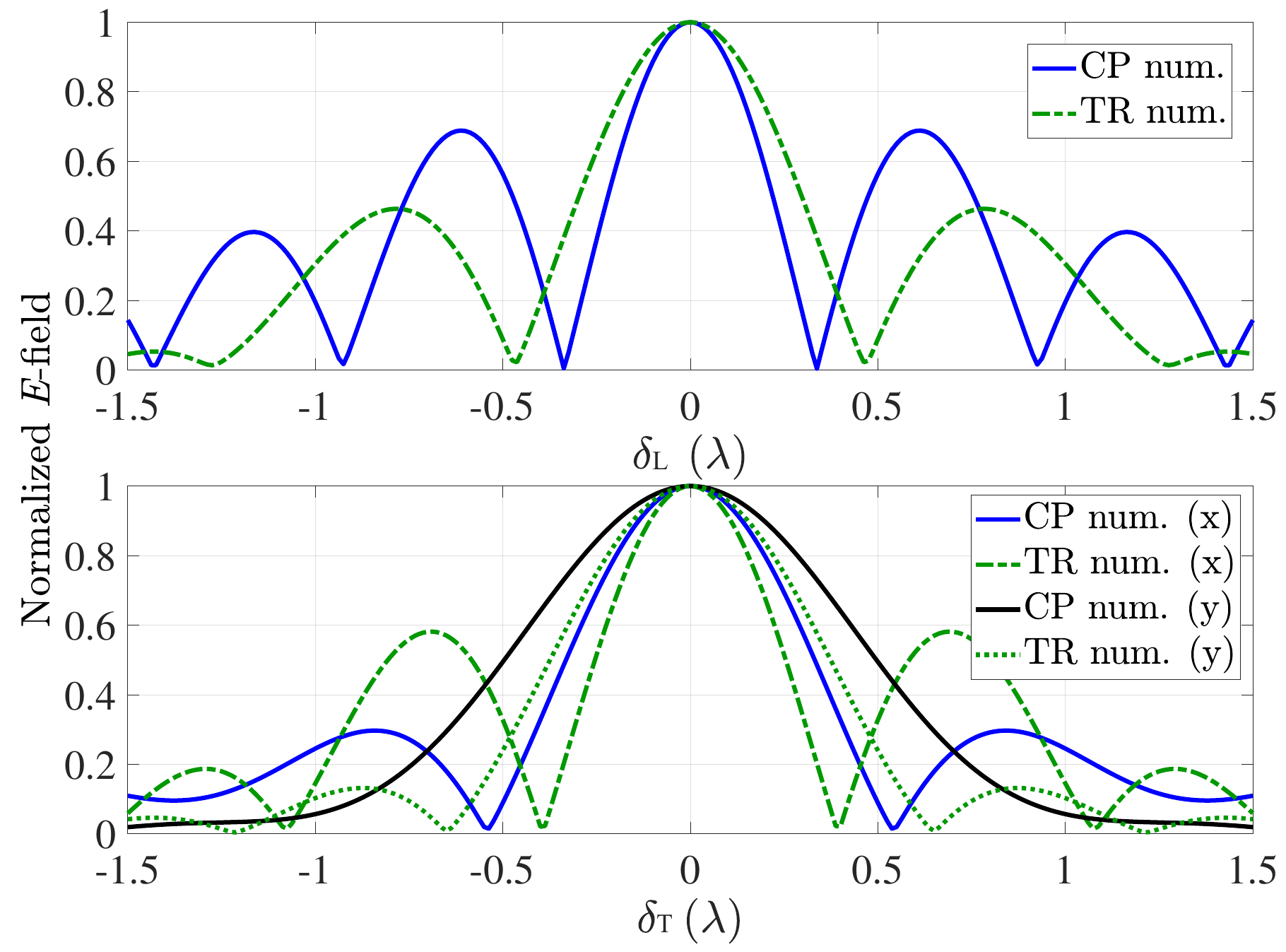} 
    \caption{Normalized electric field amplitude of $E_{\T z}$ for transverse polarization versus $\delta_{\T L}$ and $\delta_{\T T}$. 
Top: variation of $E_{\T z}$ with $\delta_{\T L}$.
Bottom: variation of $E_{\T z}$ with $\delta_{\T T}$. Peak amplitude normalized to 1.
 Unless otherwise specified, the CP method is used by default if TR is not marked in the figure.}
    \label{fig:E_Z_V}
\end{figure}

Since the co-polarized response is much stronger than the cross-polarized components in a transversely polarized array, our analysis concentrates on the resolution of $E_{\T x}^{\T v}$. For simplicity, we directly present the numerical results here. As shown in Fig.~\ref{fig:E_Z_V},
 under the CP scheme, the focal depth exhibits relatively poor performance with higher sidelobes, whereas the focal width is better controlled and the average sidelobe level is comparatively low in both directions. In contrast, the TR scheme leads to pronounced sidelobes in both the axial and transverse directions of the focal region.

\section{Electric field components for transversely polarized Hertzian dipole array}\label{app:vertical hertizian dipole}

Following the field decomposition adopted for the longitudinally  polarized Hertzian dipole, the transversely polarized Hertzian dipole can be decomposed as
\begin{multline}\label{eq:V_EX expression}
E_{\T x}^{\T v}
= \sum_{m=-M/2}^{M/2} \sum_{n=1}^{N}
\Big[
\big((m-1)d - z\big)^2  \\
+ \big( y - a \sin \phi_n \big)^2
\Big]\,
Q_{mn}(x, y, z),
\end{multline}
\begin{equation}\label{eq:V_EY expression}
E_{\T y}^{\T v} = \sum_{m=-M/2}^{M/2} \sum_{n=1}^{N} \left( x - a \cos \phi_n \right) \left( y - a \sin \phi_n \right) Q_{mn}(x, y, z),
\end{equation}
\begin{equation}\label{eq:V_EZ expression}
E_{\T z}^{\T v} = \sum_{m=-M/2}^{M/2} \sum_{n=1}^{N} \left( x - a \cos \phi_n \right) \left(d_m - z \right) Q_{mn}(x, y, z).
\end{equation}

Under the CP method, by substituting $z_{\T f}$ into \eqref{eq:V_EX expression} and performing the surface integral over both the circumferential and axial dimensions, we obtain,
\begin{multline}
\frac{4 \lambda^2 E_{\T x}^\T {C_v}(z_{\T f})}{R_{\T {e}} w_{\T {m}} a}
= \int_{-L/2}^{L/2}\!\!\int_{0}^{2\pi}
\frac{a^3\sin^2\phi + a(z_{\T f} - d_m)^2}
{\bigl[a^2 + (z_{\T f} - d_m)^2\bigr]^{3/2}}
\,\diff \phi\,\diff l \\[0.6ex]
= \pi \Bigl[
-\tfrac{L}{\sqrt{L^2 + 4L z_{\T f} + 4(a^2+z_{\T f}^2)}}
-\tfrac{L}{\sqrt{L^2 - 4L z_{\T f} + 4(a^2+z_{\T f}^2)}} \\[0.4ex]
\quad -\tfrac{2z_{\T f}}{\sqrt{L^2 + 4L z_{\T f} + 4(a^2+z_{\T f}^2)}}
+\tfrac{2z_{\T f}}{\sqrt{L^2 - 4L z_{\T f} + 4(a^2+z_{\T f}^2)}} \\[0.4ex]
\quad +2\ln\bigl(\sqrt{L^2 + 4L z + 4(a^2+z_{\T f}^2)} + L + 2z_{\T f}\bigr)\\
\quad -2\ln\bigl(\sqrt{L^2 - 4L z_{\T f} + 4(a^2+z_{\T f}^2)} - L + 2z_{\T f}\bigr)
\Bigr].
\end{multline}

Similarly, by substituting \(z_{\T f}\) into \eqref{eq:V_EY expression}, we can compute
\begin{multline}\label{eq:V_EY integration}
\frac{2\lambda^2 E_{\T y}^{\T {C_v}}(z_{\T f})}{R_{\T {e}} w_{\T {m}} a}
= \int_{-L/2}^{L/2}\!\!\int_{0}^{2\pi}
\frac{a^3\lvert \sin\phi\cos\phi \rvert}
{\bigl[a^2 + (z_{\T f} - d_m)^2\bigr]^{3/2}}
\,\diff \phi\,\diff l \\[0.6ex]
= 2z_{\T f}\Bigl(
-\tfrac{1}{\sqrt{4a^2 + (L - 2z_{\T f})^2}}
+\tfrac{1}{\sqrt{4a^2 + (L + 2z_{\T f})^2}}
\Bigr)\\[0.4ex]
\quad +L\Bigl(
\tfrac{1}{\sqrt{4a^2 + (L - 2z_{\T f})^2}}
+\tfrac{1}{\sqrt{4a^2 + (L + 2z_{\T f})^2}}
\Bigr).
\end{multline}

By substituting \(z_{\T f}\) into \eqref{eq:V_EZ expression}, we will get the following expression
\begin{multline}\label{eq:V_EZ integration}
\frac{\lambda^2 E_{\T z}^{\T {C_v}}(z_{\T f})}{R_{\T {e}} w_{\T {m}} a}
= \int_{-L/2}^{L/2}\!\!\int_{0}^{2\pi}
\frac{a^2\,\lvert\cos\phi\rvert\,\lvert z_{\T f} - d_m\rvert}
{\bigl[a^2 + (z_{\T f} - d_m)^2\bigr]^{3/2}}
\,\diff \phi\,\diff l \\[0.6ex]
= 2
-2a\Bigl[
\tfrac{1}{\sqrt{L^2 - 4Lz + 4(a^2 + z^2)}} \\
\quad
+\tfrac{1}{\sqrt{L^2 + 4Lz + 4(a^2 + z^2)}}
\Bigr].
\end{multline}

In this case, as $L \gg a$, $E_{\T y}^\T {C_{v}}$ and $E_{\T z}^\T {C_{v}}$ converge to \({w_{\T {m}} R_{\T {e}} a /{\lambda^2}}\) and \({2w_{\T {m}} R_{\T {e}} a /{\lambda^2}}\), respectively. However, the co-polarized component $E_{\T x}^\T {C_{v}}$ does not converge; instead, it grows slowly with $L$ in a logarithmic manner. Under the TR scheme, the three field components can be expressed, respectively, as
\begin{equation}
\begin{cases}
E_{\T x}^{\T {T_v}}
= F_3\!\left(\dfrac{L}{2}-z_{\T f}\right)
  - F_3\!\left(-z_{\T f}-\dfrac{L}{2}\right),\\[4pt]
\displaystyle
\frac{\lambda^2 F_3(u)}{D_{\T r} R_{\T {e}} }
= \frac{\pi}{128}\left(
41\,\tan^{-1}\!\left(\frac{u}{R}\right)
- \frac{17a^{3}u+23au^{3}}{(a^{2}+u^{2})^{2}}
\right).
\end{cases}
\end{equation}
\begin{equation}
\begin{cases}
E_{\T y}^{\T {T_v}}
= F_4\!\left(\dfrac{L}{2}-z_{\T f}\right)
  - F_4\!\left(-z_{\T f}-\dfrac{L}{2}\right),\\[4pt]
\displaystyle
\frac{\lambda^2 F_4(u)}{D_{\T r} R_{\T {e}} }
= \frac{\pi}{128}\left(
\frac{5a^{3}u+3au^{3}}{(a^{2}+u^{2})^{2}}
+3\,\tan^{-1}\!\left(\frac{u}{a}\right)
\right).
\end{cases}
\end{equation}
\begin{equation}
\begin{cases}
E_{\T z}^{\T {T_v}}
= F_5\!\left(\dfrac{L}{2}-z_{\T f}\right)
  - F_5\!\left(-z_{\T f}-\dfrac{L}{2}\right),\\[4pt]
\displaystyle
\frac{\lambda^2 F_5(u)}{D_{\T r} R_{\T {e}} }
= \frac{\pi}{32}\left(
\frac{au^{3}-a^{3}u}{(a^{2}+u^{2})^{2}}
+\tan^{-1}\!\left(\frac{u}{a}\right)
\right).
\end{cases}
\end{equation}
In the asymptotic regime where $L \gg a$, the corresponding results converge to $41\pi^{2}{D_{\T r} R_{\T {e}} /(108\lambda^2)}$, $3\pi^{2}{D_{\T r} R_{\T {e}} /(108\lambda^2)}$, $\pi^{2}{D_{\T r} R_{\T {e}}/(32\lambda^2)}$, respectively.

\bibliographystyle{IEEEtran}
\bibliography{citations}

\newpage

\vfill

\end{document}